\documentclass[a4paper,11pt]{article}
\pdfoutput=1 
\usepackage{jheppub} 

\usepackage[T1]{fontenc} 

\usepackage{bm}

\newcommand{\g}{\gamma}
\newcommand{\sig}{\sigma}

\newcommand{\open}{\sphericalangle}

\newcommand{\nn}{\nonumber}
\newcommand{\beq}{\begin{eqnarray}}
\newcommand{\eeq}{\end{eqnarray}}

\title{Realistic estimate of valence transversity distributions from inclusive dihadron production}

\author[a]{Marco Radici}
\author[b,c]{A.~Courtoy}
\author[d,a]{Alessandro Bacchetta}
\author[a]{Marco Guagnelli}

\affiliation[a]{INFN Sezione di Pavia,\\ via Bassi 6, I-27100 Pavia, Italy}
\affiliation[b]{IFPA, AGO Department, Universit\'e de Li\`ege,\\ B\^at. B5, Sart Tilman B-4000 Li\`ege, Belgium}
\affiliation[c]{Divisi\'on de Ciencias e Ingener\'ias, Universidad de Guanajuato, C.P. 37150, Le\'on, Guanajuato, M\'exico}
\affiliation[d]{Dipartimento di Fisica, Universit\`a di Pavia,\\  via Bassi 6, I-27100 Pavia, Italy}

\emailAdd{marco.radici@pv.infn.it}
\emailAdd{aurore.courtoy@ulg.ac.be}
\emailAdd{alessandro.bacchetta@unipv.it}
\emailAdd{marco.guagnelli@pv.infn.it}

\abstract{We present an updated extraction of the transversity parton distribution based on the analysis of pion-pair production in deep-inelastic scattering off transversely polarized targets in collinear factorization. Data for proton and deuteron targets make it possible to perform a flavor separation of the valence components of the transversity distribution, using di-hadron fragmentation functions taken from the semi-inclusive production of two pion pairs in back-to-back jets in $e^+ e^-$ annihilation. The  $e^+ e^-$ data from Belle have been reanalyzed using the replica method and a more realistic estimate of the uncertainties on the chiral-odd interference fragmentation function has been obtained. Then, the transversity distribution has been extracted by using the most recent and more precise COMPASS data for deep-inelastic scattering off proton targets. Our results represent the most accurate estimate of the uncertainties on the valence components of the transversity distribution currently available. }

\date{\today, \currenttime}

\begin{document}
\maketitle
\flushbottom

\section{Introduction}
\label{s:intro}

Parton distribution functions (PDFs) describe combinations of number densities of quarks and gluons in a fast-moving hadron.  At leading twist, the spin structure of spin-half hadrons is specified by three PDFs. The least known one is the chiral-odd transverse polarization distribution $h_1$ (transversity) because it can be measured only in processes with two hadrons in the initial state, or one hadron in the initial state and at least one hadron in the final state ({\it e.g.} Semi-Inclusive DIS - SIDIS).  

The transversity distribution was extracted for the first time by combining data on polarized single-hadron SIDIS together with data on almost back-to-back emission of two hadrons in $e^+ e^-$ annihilations~\cite{Anselmino:2008jk,Anselmino:2013vqa}. The difficult part of this analysis lies in the factorization framework used to interpret the data, since it involves Transverse Momentum Dependent partonic functions (TMDs). QCD evolution of TMDs must be included to analyze SIDIS and $e^+ e^-$ data obtained at very different scales, but an active debate is still ongoing about the implementation of these effects (see, e.g., Refs.~\cite{Collins:2014loa,Echevarria:2014rua,Kang:2014zza} and references therein).  

Alternatively, transversity can be extracted in the standard framework of collinear factorization using SIDIS with two hadrons detected in the final state. In this case, $h_1$ is multiplied by a specific chiral-odd Di-hadron Fragmentation Function 
(DiFF)~\cite{Collins:1994kq,Jaffe:1998hf,Radici:2001na}, which can be extracted from the corresponding $e^+ e^-$ annihilation process leading to two back-to-back hadron pairs~\cite{Boer:2003ya,Courtoy:2012ry}. In the collinear framework, evolution equations of DiFFs can be computed~\cite{Ceccopieri:2007ip}. Using $(\pi^+ \pi^-)$ SIDIS data off a transversely polarized proton target from HERMES~\cite{Airapetian:2008sk} and Belle data for the process 
$e^+ e^- \to (\pi^+ \pi^-) (\pi^+ \pi^-) X$~\cite{Vossen:2011fk}, a point-by-point extraction of transversity was performed for the first time in the collinear framework~\cite{Bacchetta:2011ip}. Later, including SIDIS data from transversely polarized proton and deuteron targets from COMPASS~\cite{Adolph:2012nw}, the valence components of up and down quarks were separated and independently parametrized~\cite{Bacchetta:2012ty}. Recently, the point-by-point extraction has been verified and extended also to the case of single-hadron SIDIS, showing that the transversity distributions obtained with two different mechanisms are compatible with each other~\cite{Martin:2014wua}. 

In this paper, we update the extraction of DiFFs from $e^+ e^-$ annihilation data by performing the fit using the replica method~\cite{Bacchetta:2012ty}. Then, using the most recent SIDIS data for charged pion pairs off a transversely polarized proton target by COMPASS~\cite{Braun:2015baa} we extract the transversity $h_1$, thus obtaining the currently most realistic estimate of the uncertainties involved. 

In Sec.~\ref{s:theory}, we summarize the theoretical framework. In Sec.~\ref{s:DiFF}, we show the results of our updated extraction of DiFFs. In Sec.~\ref{s:h1}, we comment the salient features of the re-extracted valence components of transversity. Finally, in Sec.~\ref{s:end} we draw some conclusions and mention possible extensions of our analysis.


\section{Theoretical framework for two-hadron SIDIS}
\label{s:theory}

We consider the process $\ell(k) + N(P) \to \ell(k') + H_1(P_1) + H_2(P_2) + X$, where $\ell$ denotes the incoming lepton with four-momentum $k$, $N$ the nucleon target with momentum $P$, mass $M$, and polarization $S$, $H_1$ and $H_2$ the produced unpolarized hadrons with momenta $P_1,\, P_2$ and masses $M_1, \, M_2$, respectively. We define the total $P_h = P_1 + P_2$ and relative $R = (P_1-P_2)/2$ momenta of the pair, with $P_h^2 = M_h^2 \ll Q^2=-q^2 \geq 0$ and $q = k - k'$ the space-like momentum transferred. As usual in SIDIS, we define also the following kinematic invariants
\begin{equation}
\centering
x = \frac{Q^2}{2\,P\cdot q} \; , 
\quad 
y = \frac{P \cdot q}{P \cdot k} \; ,
\label{e:invariants}
\end{equation}
\begin{equation}
\centering
z = \frac{P \cdot P_h}{P\cdot q}\equiv z_1 + z_2 \; ,
\quad
\zeta = \frac{2 R \cdot P}{P_h \cdot P} = \frac{z_1 - z_2}{z}  \; ,
 \label{e:zetaz}
\end{equation}
where $z_1, \, z_2,$ are the fractional energies carried by the two final hadrons.

\begin{figure}[htb]
  \centering
 \includegraphics[width=0.6\textwidth]{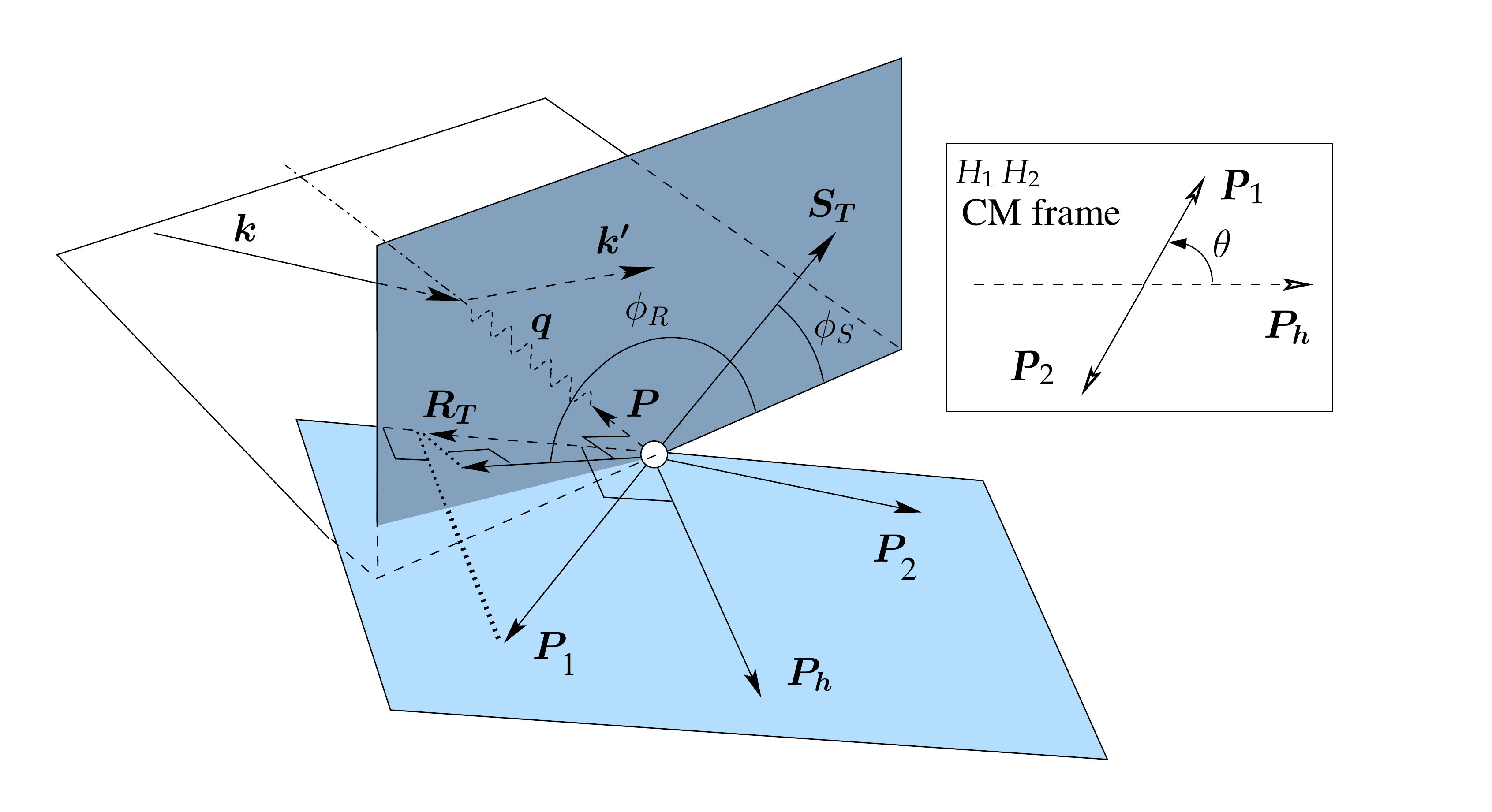}
  \caption{Kinematics of the two-hadron semi-inclusive production. The azimuthal angles $\phi_{R}$ of the component 
            $\boldsymbol{R}_T$ of the dihadron relative momentum , and  $\phi_S$ of the component $\boldsymbol{S}_T$ of 
            the target polarization, transverse to both the virtual-photon and target-nucleon momenta $\boldsymbol{q}$ and 
            $\boldsymbol{P}$, respectively, are evaluated in the virtual-photon-nucleon center-of-momentum frame. }
  \label{f:kin}
\end{figure}

The kinematics of the process is depicted in Fig.~\ref{f:kin} (see also Refs.~\cite{Airapetian:2008sk,Bacchetta:2012ty}). Of particular relevance are the azimuthal angles of the $R$ and $S$ vectors. In fact, for DiFFs it is natural to introduce the vector $R_T$  as the component of $R$ perpendicular to $P$ and $P_h$. However, the cross section will depend on the azimuthal angles of both $R_T$ and $S$ measured in the plane perpendicular to $(P, q)$. We denote the latter ones  as $\phi_R$ and 
$\phi_S$, respectively. In Ref.~\cite{Gliske:2014wba}, the covariant definition of $\phi_R$ and $\phi_S$ is derived and compared with other non-covariant definitions available in the literature, pointing out the potential differences depending on the choice of the reference frame. For the purpose of this paper, we express $\phi_R$ and $\phi_S$ in the target rest frame:
\begin{align}
\phi_{R} \equiv 
&\frac{(\bm{q} \times \bm{k}) \cdot \bm{R}_T}{\vert (\bm{q} \times \bm{k}) \cdot \bm{R}_T\vert}  
\arccos {
\frac{(\bm{q} \times \bm{k}) \cdot (\bm{q}\times \bm{R}_T)}
        {\vert \bm{q} \times \bm{k}\vert \vert \bm{q}\times \bm{R}_T\vert }   }
\; , \nn \\ 
\phi_{S} \equiv
 &\frac{(\bm{q}\times \bm{k}) \cdot \bm{S}_T}{\vert (\bm{q}\times \bm{k}) \cdot \bm{S}_T\vert} 
 \arccos {
 \frac{(\bm{q} \times \bm{k}) \cdot (\bm{q} \times \bm{S}_T)}
         {\vert \bm{q}\times \bm{k}\vert \vert \bm{q} \times \bm{S}_T\vert }   }
\; . \label{e:angles}
\end{align}
Equation~(\ref{e:angles}) is valid also in any frame reached from the target rest frame by a boost along $\bm q$, up to corrections of order $O(1/Q^2)$. 

We also define the polar angle $\theta$ which is the angle between the direction of the back-to-back emission in the center-of-mass (cm) frame of the two final hadrons, and the direction of $P_h$ in the photon-proton cm frame (see Fig.~\ref{f:kin}). We have
\begin{align}
\vert \bm{R} \vert &= \frac{1}{2}\, \sqrt{M_h^2 - 2(M_1^2+M_2^2) + (M_1^2-M_2^2)^2/M_h^2} 
\; ,  \nn \\
\bm{R}_T &= \bm{R} \sin \theta  \; . 
\label{e:Rvect}
\end{align}
The invariant $\zeta$ of Eq.~(\ref{e:zetaz}) can be shown to be a linear polynomial in $\cos \theta$~\cite{Bacchetta:2002ux}.

In the one-photon exchange approximation and neglecting the lepton mass, to leading order in the couplings, the differential cross section for the two-hadron SIDIS of an unpolarized lepton off a transversely polarized nucleon target 
reads~\cite{Bacchetta:2012ty}
\begin{align}
\lefteqn{\frac{d\sig}{dx \, dy\, d\psi \,dz\, d\phi_R\, d M_{h}^2\,d \cos{\theta}} =}  \nn \\ 
& \quad
\frac{\alpha^2}{x y\, Q^2}\, \Biggl\{ A(y) \, F_{UU}  + |\bm{S}_T|\, B(y) \, \sin(\phi_R+\phi_S)\,  F_{UT}^{\sin (\phi_R +\phi_S)}
\Biggr\} \; ,
\label{e:crossmaster}
\end{align}
where $\alpha$ is the fine structure constant, $A(y) = 1 - y + y^2/2$, $B(y) = 1 - y$, and the angle $\psi$ is the azimuthal angle of $k'$ around the lepton beam axis with respect to the direction of $S$. In DIS kinematics, it turns out 
$d\psi \approx d\phi_S$~\cite{Diehl:2005pc}. 

In the limit $M_h^2 \ll Q^2$, the structure functions in Eq.~(\ref{e:crossmaster}) can be written as products of PDFs and 
DiFFs~\cite{Bianconi:1999cd,Radici:2001na,Bacchetta:2002ux}: 
\begin{align} 
F_{UU} & = x \sum_q e_q^2\, f_1^q(x; Q^2)\, D_1^q\bigl(z,\cos \theta, M_h; Q^2\bigr) \; , 
\label{e:FUU} \\
F_{UT}^{\sin (\phi_R +\phi_S)} &=  \frac{|\bm R| \sin \theta}{M_h}\, x\, 
\sum_q e_q^2\,  h_1^q(x; Q^2)\,H_1^{\open\, q}\bigl(z,\cos \theta, M_h; Q^2\bigr) \; , 
\label{e:FUT}
\end{align}
where $e_q$ is the fractional charge of a parton with flavor $q$. The $D_1^q$ is the DiFF describing the hadronization of an  unpolarized parton with flavor $q$ into an unpolarized hadron pair. The $H_1^{\open\, q}$ is a chiral-odd DiFF describing the correlation between the transverse polarization of the fragmenting parton with flavor $q$ and the azimuthal orientation of the plane containing the momenta of the detected hadron pair. 

Since $M_h^2 \ll Q^2$, the hadron pair can be assumed to be produced mainly in relative $s$ or $p$ waves, suggesting that the DiFFs can be conveniently expanded in partial waves. From Eq.~(\ref{e:Rvect}) and from the simple relation between 
$\zeta$ and $\cos \theta$, DiFFs can be expanded in Legendre polynomials in $\cos \theta$~\cite{Bacchetta:2002ux}. After averaging over $\cos \theta$, only the term corresponding to the unpolarized pair being created in a relative $\Delta L=0$ state survives in the $D_1$ expansion, while the interference with $|\Delta L| = 1$ survives for 
$H_1^{\open}$~\cite{Bacchetta:2002ux}. The simplification holds even if the $\theta$ dependence in the acceptance is not complete but symmetric about $\theta = \pi / 2$. Without ambiguity, the two surviving terms will be identified with $D_1$ and $H_1^\open$, respectively. 

By inserting the structure functions of Eqs.~(\ref{e:FUU}), (\ref{e:FUT}) into the cross section~(\ref{e:crossmaster}), we can define the single-spin asymmetry (SSA)~\cite{Radici:2001na,Bacchetta:2002ux,Bacchetta:2006un}
\begin{equation}
A_{\mathrm{SIDIS}} (x, z, M_h; Q) = - \frac{B(y)}{A(y)} \,\frac{|\bm{R} |}{M_h} \, 
\frac{ \sum_q\, e_q^2\, h_1^q(x; Q^2)\, H_1^{\open\, q}(z, M_h; Q^2)    } 
        { \sum_q\, e_q^2\, f_1^q(x; Q^2)\, D_{1}^q (z, M_h; Q^2) } \; .
\label{e:ssa}
\end{equation} 

For the specific case of $\pi^+ \pi^-$ production, isospin symmetry and charge conjugation suggest $D_1^q = D_1^{\bar{q}}$ and $H_1^{\open\, q} = - H_1^{\open\, \bar{q}}$ for $q = u,d,s,$ and also $H_1^{\open\, u} = - H_1^{\open\, d} $ and $H_1^{\open\, s} = 0$~\cite{Bacchetta:2006un,Bacchetta:2011ip,Bacchetta:2012ty}. Moreover, from 
Eq.~(\ref{e:ssa}) the $x$-dependence of transversity is more conveniently studied by integrating the $z$- and 
$M_h$-dependences of DiFFs. So, in the analysis the actual combinations used for the proton 
target are~\cite{Bacchetta:2012ty} 
\begin{equation} 
\begin{split}
x\, h_1^{p} &(x; Q^2) \equiv x \, h_1^{u_v}(x; Q^2) - {\textstyle \frac{1}{4}}\, x h_1^{d_v}(x; Q^2) \\
&= -\frac{ A^p_{\text{SIDIS}} (x; Q^2)  }{n_u^{\uparrow}(Q^2)} \, \frac{A(y)}{B(y)}\, \frac{9}{4}\, 
\sum_{q=u,d,s} \, e_q^2 \, n_q (Q^2)\, x f_1^{q+\bar{q}}(x; Q^2)  \; ,
\label{e:h1p} 
\end{split}   
\end{equation} 
and for the deuteron target are 
\begin{equation} 
\begin{split} 
 x\, h_1^{D} &(x; Q^2) \equiv x \, h_1^{u_v}(x; Q^2)+ x h_1^{d_v}(x; Q^2)   \\
 &=- \frac{A^D_{\text{SIDIS}}(x; Q^2)}{n_u^{\uparrow}(Q^2)} \, 3  \, 
 \sum_{q=u,d,s} \, \left[ e_q^2 \, n_q (Q^2) + e^2_{\tilde{q}} \, n_{\tilde{q}} (Q^2) \right] \, x f_1^{q+\bar{q}}(x; Q^2) \; ,
\label{e:h1D}
\end{split} 
\end{equation} 
where $h_1^{q_v} \equiv h_1^q - h_1^{\bar{q}}$, $f_1^{q+\bar{q}} \equiv f_1^q + f_1^{\bar{q}}$, $\tilde{q} = d,u,s$ if 
$q = u,d,s,$ respectively ({\it i.e.} it reflects isospin symmetry of strong interactions inside the deuteron), and 
\begin{align} 
n_q(Q^2) &= \int dz \int dM_h \, D_1^q (z, M_h; Q^2)  \; , \label{e:nq}  \\
n_q^\uparrow (Q^2) &= \int dz \int dM_h \, \frac{|\bm{R}|}{M_h}\, H_1^{\open\, q}(z,M_h; Q^2) \; .
\label{e:nqperp}
\end{align}

Using Eqs.~(\ref{e:h1p}) and (\ref{e:h1D}), we can extract the valence components of transversity from the measurement of SSA $A^p_{\text{SIDIS}}$ and $A^D_{\text{SIDIS}}$, and from the knowledge of DiFFs through Eqs.~(\ref{e:nq}) and (\ref{e:nqperp}). 


\section{Extraction of Di-hadron Fragmentation Functions}
\label{s:DiFF}

The unknown DiFFs in Eqs.~(\ref{e:nq}) and~(\ref{e:nqperp}) can be extracted from the process 
$e^+ e^- \to (\pi^+ \pi^-)_{\text{jet}} (\pi^+ \pi^-)_{\overline{\text{jet}}} X$. Namely, an electron and a positron annihilate producing a virtual photon (whose time-like momentum defines the hard scale $Q^2 \geq 0$). Then, the photon decays in a quark and an antiquark, each one fragmenting into a residual jet. The two jets are produced in a back-to-back configuration; this is granted by requiring that the total momentum $P_h$ of the $(\pi^+ \pi^-)_{\text{jet}}$ pair in the quark jet and the total momentum $\bar{P}_h$ of the $(\pi^+ \pi^-)_{\overline{\text{jet}}}$ pair in the antiquark jet are such that 
$P_h \cdot \bar{P}_h \approx Q^2$ (in the following, all overlined variables will refer to the antiquark jet). 

The leading-twist cross section in collinear factorization, namely by integrating upon all transverse momenta but ${\bf R}_T$ and ${\bf \bar{R}}_T$, can be written as~\cite{Courtoy:2012ry}
\begin{equation}
\frac{d\sigma}{d\cos\theta_2 dz d\cos\theta dM_h d\phi_R d\bar{z} d\cos\bar{\theta} d\bar{M}_h d\bar{\phi}_R} = 
\frac{1}{4\pi^2}\, d\sigma^0 \, \bigg( 1+ \cos (\phi_R + \phi_{\bar{R}} ) \, A_{e^+e^-} \bigg) \; , 
\label{e:e+e-cross}
\end{equation}
where $\theta_2$ is the angle in the lepton plane formed by the positron direction and $\bm{P}_h$ (according to the Trento 
conventions~\cite{Bacchetta:2004jz}), and the azimuthal angles $\phi_R$ and $\phi_{\bar{R}}$ give the orientation of the planes containing the momenta of the pion pairs with respect to the lepton plane (see Fig.1 of Ref.~\cite{Courtoy:2012ry} for more details). The $d\sigma^0$ is the unpolarized cross section producing an azimuthally flat distribution of pion pairs coming from the fragmentation of unpolarized quarks. The term $A_{e^+e^-}$ represents the so-called Artru-Collins asymmetry and is given 
by~\cite{Boer:2003ya}
\begin{eqnarray}
A_{e^+e^-} &= &\frac{\sin^2 \theta_2}{1+ \cos^2 \theta_2} \, \sin \theta \, \sin \bar{\theta} \, \frac{|\bm{R}|}{M_h} \, \frac{|\bar{\bm{R}}|}{\bar{M}_h} \, \frac{\sum_q e_q^2\, H_1^{\sphericalangle\, q}(z,M_h; Q^2)\, H_1^{\sphericalangle\, \bar{q}}(\bar{z},\bar{M}_h; Q^2)}
        {\sum_q e_q^2\, D_1^q(z,M_h; Q^2)\, D_1^{\bar{q}}(\bar{z},\bar{M}_h; Q^2)} \, .
\label{e:e+e-ssa}
\end{eqnarray} 

The $H_1^{\sphericalangle\, q}$ can be extracted from the Artru-Collins asymmetry by conveniently integrating upon the  hemisphere of the antiquark jet. For $(\pi^+ \pi^-)$ production, isospin and charge conjugation symmetries of DiFFs imply $n_{\bar{q}} (Q^2) = n_q (Q^2)$ for $q=u,d,s,c,$ and $n^\uparrow_{\bar{q}} (Q^2) = - n^\uparrow_q (Q^2)$ for $q=u,d,$ neglecting other components and with the further constraint $n^\uparrow_{u} (Q^2) = - n^\uparrow_d (Q^2)$. Then, 
Eq.~(\ref{e:e+e-ssa}) is simplified to~\cite{Courtoy:2012ry}
\begin{equation}
A_{e^+e^-} = - \frac{\sin^2 \theta_2}{1+ \cos^2 \theta_2} \, \sin \theta \, \sin \bar{\theta} \, \frac{5}{9}\, 
\frac{H (z,M_h; Q^2)}{D (z,M_h; Q^2)} \, , 
\label{e:e+e-ssafit}
\end{equation} 
where 
\beq
D(z, M_h; Q^2) &= &\frac{4}{9}\, D_1^u (z, M_h; Q^2) \, n_u (Q^2) + \frac{1}{9}\, D_1^d (z, M_h; Q^2) \, n_d (Q^2) \nn \\
&+ &\frac{1}{9}\, D_1^s (z, M_h; Q^2) \, n_s (Q^2) + \frac{4}{9}\, D_1^c (z, M_h; Q^2) \, n_c (Q^2) \; , 
\label{e:ssafitden}
\eeq
and
\begin{eqnarray} 
H(z, M_h; Q^2) &\equiv &\frac{|\bm{R}|}{M_h} \, H_{1}^{\sphericalangle u}(z, M_h; Q^2)\, n_u^\uparrow (Q^2) \nonumber \\
&= &- \frac{1+\cos^2 \theta_2}{\sin^2 \theta_2}\, \frac{9}{5} \,\frac{1}{\sin\theta \, \sin\bar{\theta}} \, D(z, M_h; Q^2) \, 
A_{e^+e^-} \; ,
\label{e:ssafitnum}
\end{eqnarray} 
with the normalization
\begin{equation}
\int dz \int dM_h \, H(z, M_h, Q^2) = [ n_u^\uparrow (Q^2)]^2 \; . 
\label{e:Hnorm}
\end{equation}

Since a measurement of the unpolarized differential cross section is still missing, the unpolarized DiFF $D_1$ is taken from our previous analysis in Ref.~\cite{Courtoy:2012ry}, where it was parametrized to reproduce the two-pion yield of the 
{\tt PYTHIA} event generator tuned to the Belle kinematics. The fitting expression at the starting scale $Q_0^2=1$ GeV$^2$ was inspired by previous model calculations~\cite{Bacchetta:2006un,Radici:2001na,Bianconi:1999uc,Bacchetta:2008wb} and it contains three resonant channels (pion pair produced by $\rho$, $\omega$, and $K^0_S$ decays) and a continuum. For each channel and for each flavor $q= u,d,s,c$, a grid of data in $(z,M_h)$ was produced using {\tt PYTHIA} for a total amount of approximately 32000 bins. Each grid was separately fitted using the corresponding parametrization of $D_1$ and evolving it to the Belle scale at $Q^2=100$ GeV$^2$. An average $\chi^2$ per degree of freedom ($\chi^2$/d.o.f.) of 1.62 was reached using in total 79 parameters. More details can be found in Ref.~\cite{Courtoy:2012ry}. 

As for the polarized DiFF, we deduce the experimental value of Eq.~(\ref{e:ssafitnum}) in each bin, denoted $H^{\text{exp}}$, by using the experimental data for the Artru-Collins asymmetry $A_{e^+e^-}$ and the corresponding average values of the angles $\theta_2, \, \theta, \, \bar{\theta},$ taken from Ref.~\cite{Vossen:2011fk}. The function $D$ is calculated from 
Eq.~(\ref{e:ssafitden}) using the unpolarized DiFFs $D_1^q$ resulting from the fit of the {\tt PYTHIA}'s two-pion yield. The experimental data for $A_{e^+e^-}$ are organized in a $(z,M_h)$ grid of 64 bins~\cite{Vossen:2011fk}. Some of them are empty or scarcely populated; therefore, we used only 46 of them~\cite{Courtoy:2012ry}. 

The fitting value in each bin, denoted $H^{\text{th}}$, is based on the following expression at the starting scale $Q_0^2=1$ 
GeV$^2$~\cite{Courtoy:2012ry}:
\begin{eqnarray}
H(z, M_h, Q_0^2; \{p\}) &= &N_H \,  2 |\bm{R}|\, (1-z) \, \exp [\g_1 (z - \g_2 M_h)] \, \text{BW} \left( m_\rho, \frac{\eta}{m_\rho} ; M_h \right) \nn \\
&\times &\Bigg[ P(0, 1, \delta_1, 0, 0 ; z) + z P(0, 0, \delta_2, \delta_3, 0; M_h) + \frac{1}{z} \, P(0, 0, \delta_4, \delta_5, 0; M_h) \Bigg] \, ,  \nn \\
& &  
\label{e:Hfit}
\end{eqnarray}
where $\{ p \}$ denotes the vector of 9 parameters 
$\{ p \} = (N_H, \g_1, \g_2, \delta_1, \delta_2, \delta_3, \delta_4, \delta_5, \eta)$. The polynomial $P$ and the Breit--Wigner function BW are defined by
\begin{eqnarray}
P (a_1, a_2, a_3, a_4, a_5; x) &= &a_1 \frac{1}{x} + a_2 + a_3 x + a_4 x^2 + a_5 x^3  \; , \nn \\
\text{BW} (m, \Gamma; x) &= &\frac{1}{(x^2-m^2)^2 + m^2 \Gamma^2}   \; .  
\label{e:P-BW}
\end{eqnarray}
The function \text{BW} is proportional to the modulus squared of a relativistic Breit--Wigner for the considered  resonant channel, and it depends on its mass $m$ and width $\Gamma$. In our case, the $\rho \to (\pi^+ \pi^-)$ decay involves the fixed parameters $m_\rho = 0.776$ GeV and $\Gamma_\rho = 0.150$ GeV. 

The function $H$ of Eq.~(\ref{e:Hfit}) is then evolved to the Belle scale using the {\tt HOPPET} code~\cite{Salam:2008qg} suitable extended to include chiral-odd splitting functions at leading order~\cite{Courtoy:2012ry}. We have used two different values of $\alpha_s (M_Z^2)$ in the evolution code, namely $0.125$~\cite{Gluck:1998xa} and $0.139$~\cite{Martin:2009iq}, in order to account for the theoretical uncertainties on the determination of the $\Lambda_{\text{QCD}}$ parameter. At variance with Ref.~\cite{Courtoy:2012ry}, the error analysis is carried out using the same Monte Carlo approach adopted in our previous extraction of transversity~\cite{Bacchetta:2012ty}. The approach consists in creating $N$ replicas of the data points. In each replica (denoted by the index $r$), the data point in the bin $(z_i,M_{h\, j})$  is perturbated by a Gaussian noise with the same variance as the experimental measurement. Each replica, therefore, represents a possible outcome of an independent measurement; for the bin $(z_i,M_{h\, j})$, we denote it by $H_{ij, r}^{\text{exp}}$. The number of replicas is chosen $N = 100$ in order to accurately reproduce the mean and standard deviation of the original data points. The standard minimization procedure is applied to each replica $r$ separately, by minimizing the following error function
\begin{equation}
E_r^2(\{p\}) = \sum_{ij}  \frac{\left[ H_{ij}^{\text{th}} (\{p\})- H_{ij,\, r}^{\text{exp}} \right]^2}{\sig_{ij}^2} \; , 
\label{e:chi2A}
\end{equation}
where $H_{ij}^{\text{th}}$ is the fitting value of Eq.~(\ref{e:Hfit}) depending on the vector $\{ p \}$ of 9 parameters, and the error $\sig_{ij}$ for each replica is taken to be equal to the error on the original data point for the bin $(z_i,M_{h\, j})$. In fact, in the expression of $H$ from Eq.~(\ref{e:ssafitnum}) the dominant source of uncertainty comes from the experimental error on the measurement of $A_{e^+e^-}$. The very large statistics available from {\tt PYTHIA} in the Monte Carlo simulation of two-pion yields makes the statistical uncertainty on $D$ negligible~\cite{Courtoy:2012ry}. However, there is still a source of systematic error that is not taken into account in this analysis. This problem can be overcome only when real data for the unpolarized cross section will become available. Meanwhile, in this work $\sigma_{ij}$ is obtained by summing in quadrature the statistical and systematic errors for the measurement of $A_{e^+e^-}$ reported by the Belle collaboration~\cite{Vossen:2011fk}, multiplied by all factors relating $A_{e^+e^-}$ to $H$ according to Eq.~(\ref{e:ssafitnum}). 

\begin{figure}
\centering
\includegraphics[width=7cm]{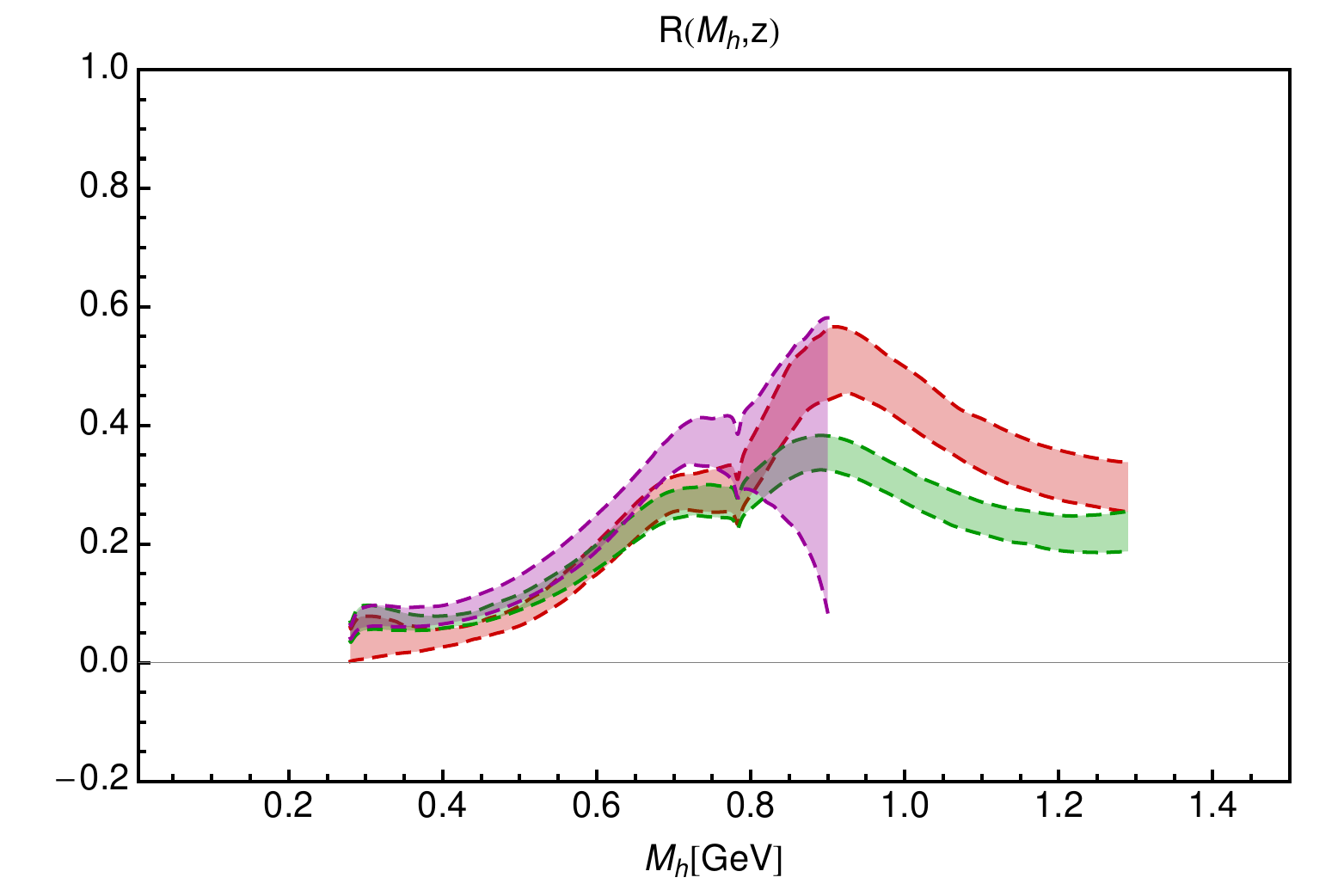} \hspace{0.5cm} \includegraphics[width=7cm]{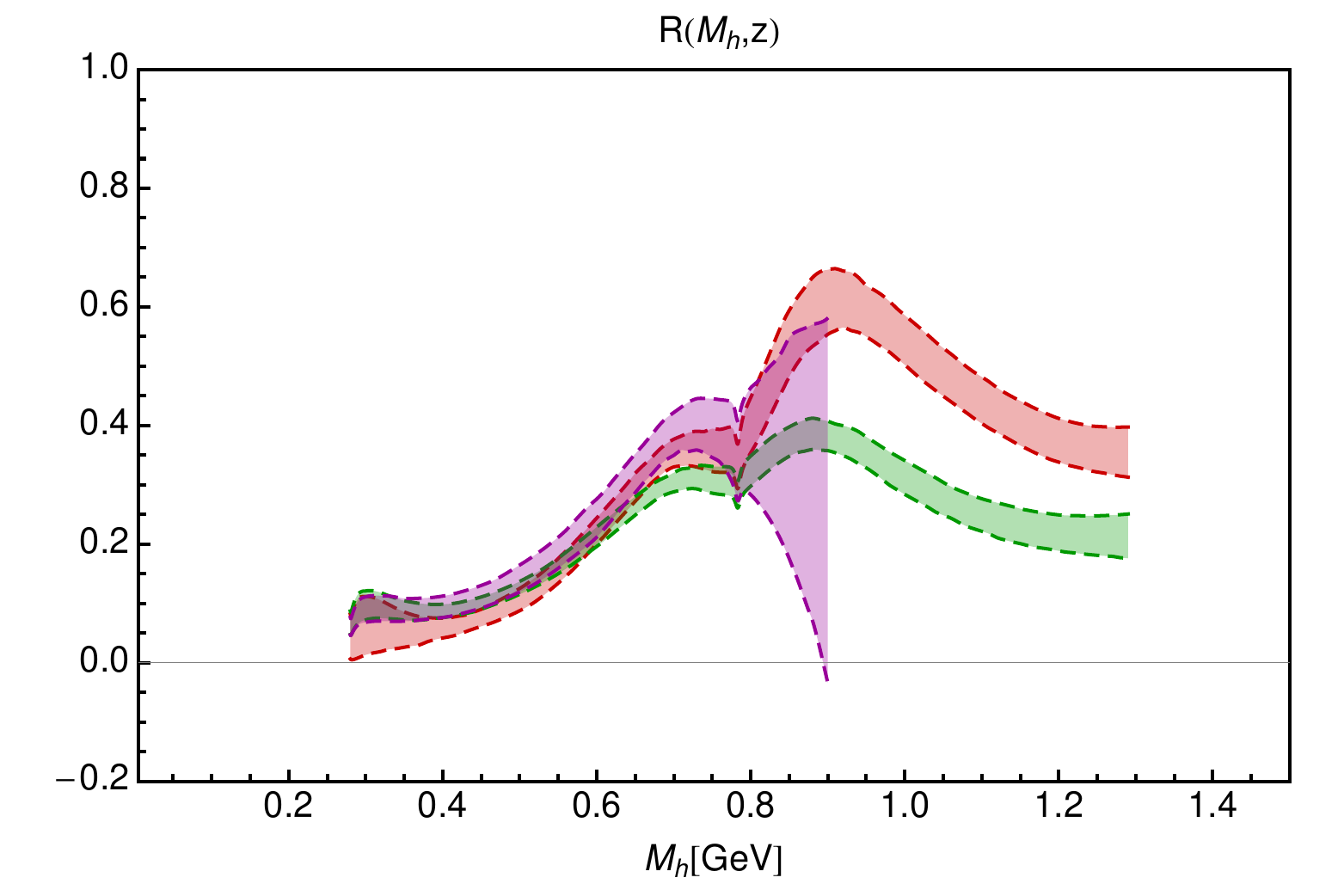} 
\caption{The ratio $R(z, M_h)$ of Eq.~(\ref{e:R}) as a function of $M_h$ at $Q_0^2=1$ GeV$^2$ for three different $z=0.25$ (shortest band), $z=0.45$ (lower band at $M_h \sim 1.2$ GeV), and $z=0.65$ (upper band at $M_h \sim 1.2$ GeV). Left panel for results obtained with $\alpha_s (M_Z^2) = 0.125$, right panel with $\alpha_s (M_Z^2) = 0.139$. For the calculation of the uncertainty bands see details in the text.} 
\label{fig:H1Mh}
\end{figure}

The  vector of parameters that initializes the minimization in Eq.~(\ref{e:chi2A}) corresponds to the one that produced the best fit in the previous analysis of Ref.~\cite{Courtoy:2012ry} using the standard Hessian method. Then, the minimization results in $N$ different vectors of best-fit parameters, $\{ p_{0r}\},\; r=1,\ldots N$. These vectors can be used to produce $N$ different values for any theoretical quantity. Using Eqs.~(\ref{e:ssafitnum}) and~(\ref{e:Hnorm}), we can produce $N$ different replicas of the polarized DiFF $H_1^{\sphericalangle u}$.  The $N$ theoretical outcomes can have any distribution, not necessarily Gaussian. Hence, the $1 \sig$ confidence interval is in general different from the 68\% interval which, in our case, can simply be obtained by rejecting for each experimental point $(z_i, M_{h\, j})$ the largest and the lowest 16\% of the $N$ values. This approach produces a more realistic estimate of the statistical uncertainty on DiFFs. In fact, we noticed that the minimization often pushes the theoretical functions towards their upper or lower bounds, where the $\chi^2$ does no longer display a quadratic dependence upon the parameters. Instead, the Monte Carlo approach does not rely on the prerequisites for the standard Hessian method to be valid. Although the minimization is performed on the function defined in Eq.~(\ref{e:chi2A}), the agreement of the $N$ theoretical outcomes with the original Belle data is better expressed in terms of the standard $\chi^2$ function~\cite{Ball:2008by}. The $\chi^2$ can be  obtained by replacing $H_{ij,\, r}^{\text{exp}}$ in Eq.~(\ref{e:chi2A}) with the corresponding value inferred from the original data set without Gaussian noise. 

\begin{figure}
\centering
\includegraphics[width=7cm]{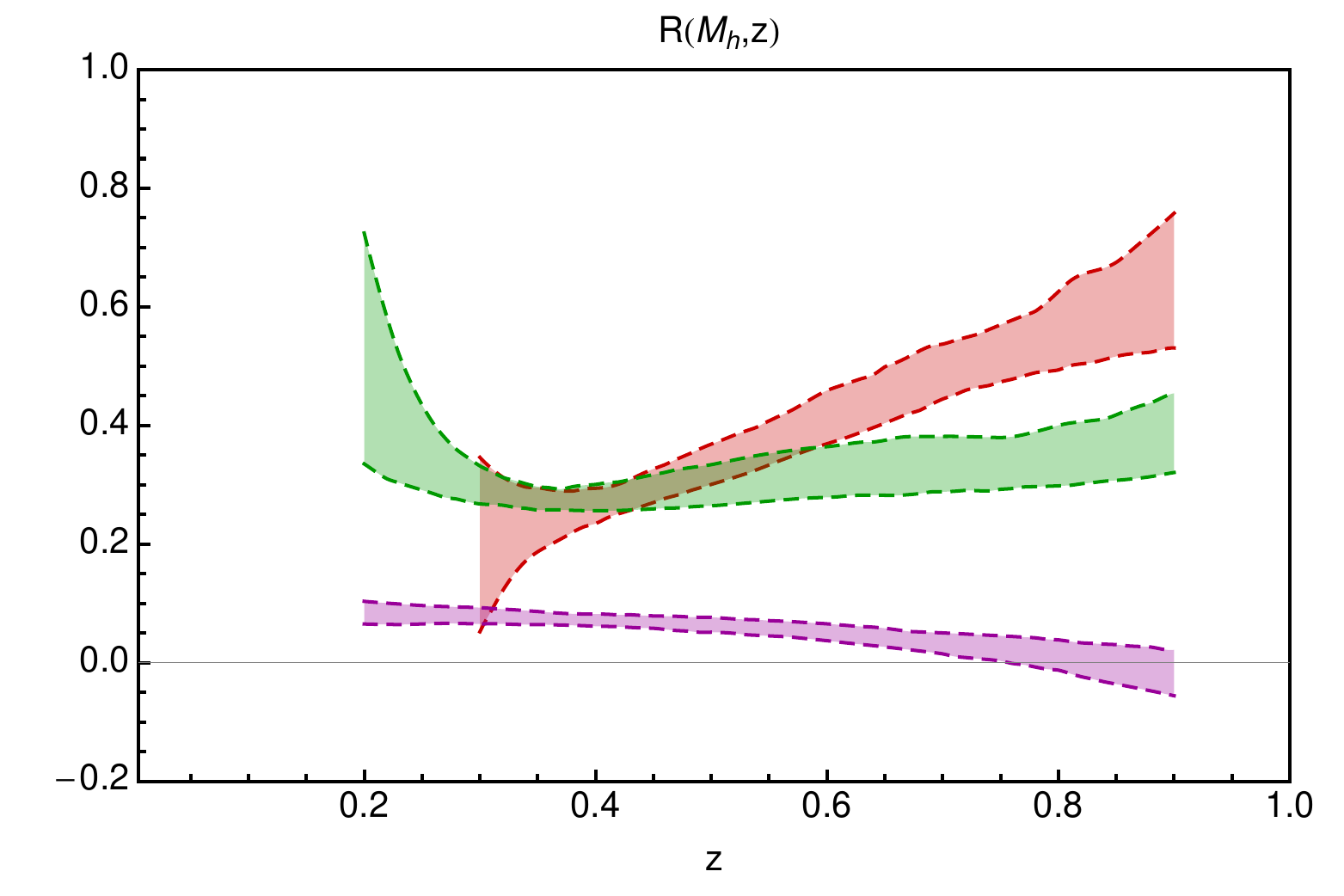} \hspace{0.5cm} \includegraphics[width=7cm]{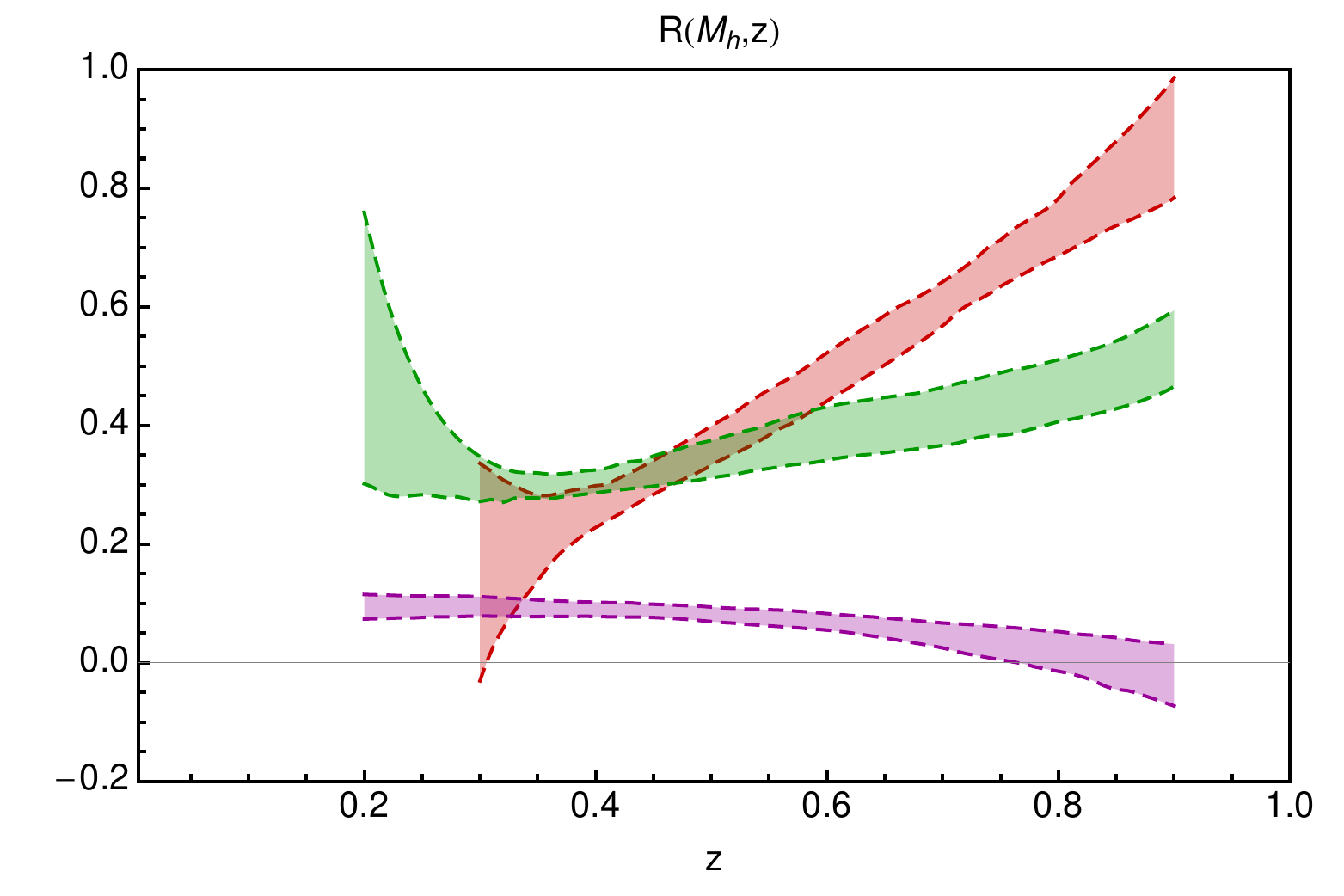}
\caption{The ratio $R(z, M_h)$ of Eq.~(\ref{e:R}) as a function of  $z$ at $Q_0^2=1$ GeV$^2$ for three different $M_h=0.4$ GeV (lower band at $z\sim 0.8$), $M_h=0.8$ GeV (mid band at $z\sim 0.8$), and $M_h=1$ GeV (upper band at $z\sim 0.8$). Left panel for results obtained with $\alpha_s (M_Z^2) = 0.125$, right panel with $\alpha_s (M_Z^2) = 0.139$. For the calculation of the uncertainty bands see details in the text.}
\label{fig:H1z}
\end{figure}

We show our results through the following ratio:
\begin{equation}
R(z, M_h) = \frac{|\bm{R}|}{M_h}\, \frac{H_1^{\open\, u} (z, M_h; Q_0^2)}{D_1^u (z, M_h; Q_0^2)} \; , 
\label{e:R}
\end{equation}
where both DiFFs are summed over all fragmentation channels and the ratio is evaluated at the hadronic scale $Q_0^2=1$ GeV$^2$. In Fig.~\ref{fig:H1Mh}, we consider the ratio $R$ as a function of the invariant mass $M_h$ for three different values of the fractional energy $z$: $z=0.25$ (shortest band), $z=0.45$ (lower band at $M_h \sim 1.2$ GeV), and $z=0.65$ (upper band at $M_h \sim 1.2$ GeV). The left panel displays the results with $\alpha_s (M_Z^2) = 0.125$, the right one with 
$\alpha_s (M_Z^2) = 0.139$. Each band corresponds to the $68\%$ of all $N = 100$ replicas, produced by rejecting the largest and lowest $16\%$ of the replicas' values for each $(z, M_h)$ point. The shortest band (for $z=0.25$) stops around 
$M_h \sim 0.9$ GeV because there are no experimental data at higher invariant masses for such low values of $z$. In this kinematic range, the fit is much less constrained and, consequently, the uncertainty band becomes larger. Comparing the two panels reveals a mild sensitivity of the results to the choice of $\alpha_s (M_Z^2)$, hence of $\Lambda_{\text{QCD}}$. Figure~\ref{fig:H1Mh} represents an update of the upper panel of Fig.~6 in Ref.~\cite{Courtoy:2012ry}, with a more realistic estimate of the statistical uncertainty on the polarized DiFF $H_1^{\open}$. 

In Fig.~\ref{fig:H1z}, the same quantity $R$ of Eq.~(\ref{e:R}) is plotted as a function of $z$ for three different values of the invariant mass: $M_h=0.4$ GeV (lower band at $z\sim 0.8$), $M_h=0.8$ GeV (mid band at $z\sim 0.8$), and $M_h=1$ GeV (upper band at $z\sim 0.8$). Again, the left panel displays the results with $\alpha_s (M_Z^2) = 0.125$ while the right one with $\alpha_s (M_Z^2) = 0.139$. The results of the two panels are similar, as we found in the previous figure, except for the bands with larger $M_h$ at very high values of $z$. Again, Fig.~\ref{fig:H1z} represents a realistic update of the bottom panel of Fig.~6 in Ref.~\cite{Courtoy:2012ry}. 

\begin{figure}
\centering
\includegraphics[width=6.5cm]{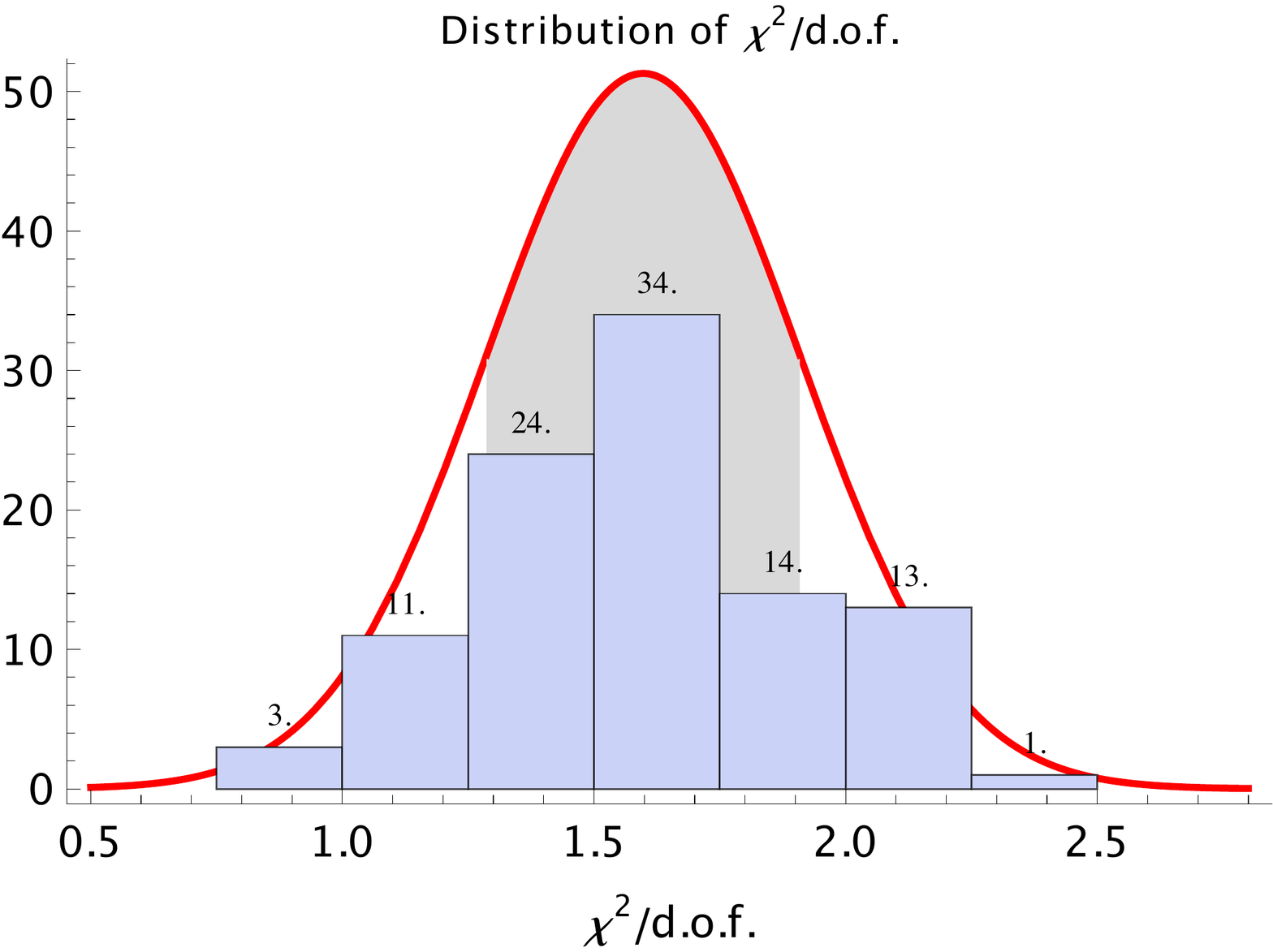} \hspace{0.5cm} \includegraphics[width=6.5cm]{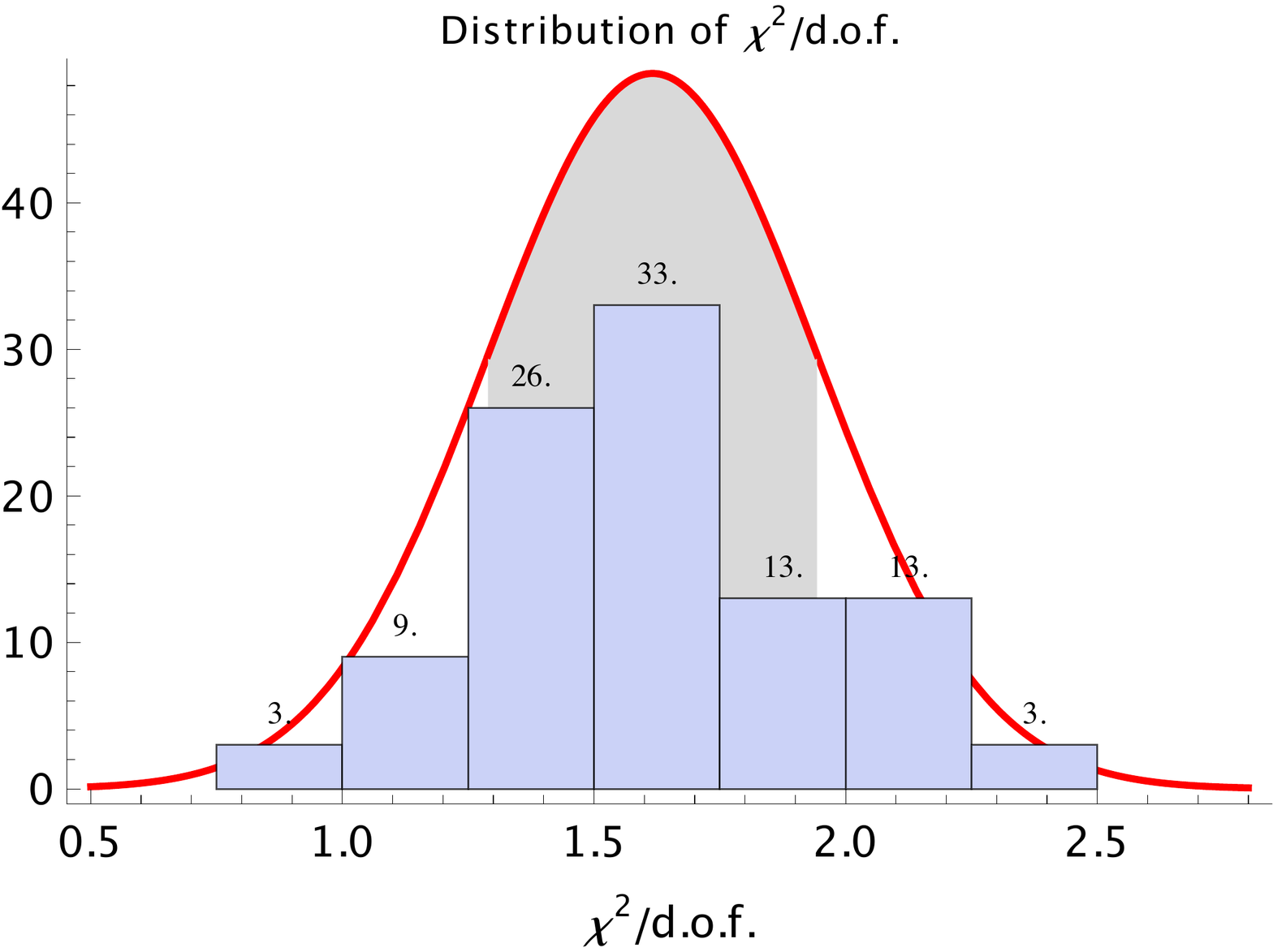}
\caption{Histogram of the distribution of $N=100$ $\chi^2$/d.o.f. for $\alpha_s (M_Z^2)=0.125$ (left panel) and 
$\alpha_s (M_Z^2)=0.139$ (right panel). The solid curve corresponds to a Gaussian distribution centered on the average of the $N=100$ $\chi^2$ values. The shaded area represents the $1\sig$ variance. The normalization of the Gaussian distribution is adapted to the histogram profile.}
\label{fig:chi2}
\end{figure}

From the minimization in Eq.~(\ref{e:chi2A}), we obtain $N$ different $\chi^2$, each one corresponding to a different vector of fitting parameters $\{p\}_r, \, r=1\dots N$. If the model is able to give a good description of the data, the distribution of the 
$N$ values of $\chi^2$/d.o.f. should be peaked at around one. In real situations, the rigidity of the model shifts the position of the peak to higher values. In Fig.~\ref{fig:chi2}, we show the histogram for the distribution of the $N$ values of the $\chi^2$/d.o.f. It is not peaked at 1 but slightly above 1.5. For sake of illustration, we compare it with the solid line representing a Gaussian distribution centered at the average value of the $N$ different $\chi^2$/d.o.f. The shaded area represents its $1\sig$ variance. As in previous figures, the left panel shows the $\chi^2$/d.o.f. obtained using $\alpha_s (M_Z^2) = 0.125$ in the evolution code, while in the right panel $\alpha_s (M_Z^2) = 0.139$. The salient features of the $\chi^2$ distribution remain the same in both cases. 

\begin{table}[h]
\begin{center}
\begin{tabular}{ |c||c|c||c|c| }
  \hline
  \multicolumn{1}{|c||}{}& \multicolumn{2}{|c||}{$\alpha_s(M_Z^2)=0.125$}& \multicolumn{2}{|c|}{$\alpha_s(M_Z^2)=0.139$} \\
  \hline
 $ \{ p \}$		&   $\{ \langle p \rangle \}$	&  $\{ \sig_p \}$ & $\{ \langle p \rangle \}$	& $\{ \sig_p \}$ \\
  \hline
  \hline	
 $N$			&	0.016 			& $0.008$ 	&  $0.015$			&	$0.007$  \\
\hline
$\g_1$		&    $-2.73$			& $0.91$ 		&  $-2.27$				&	$0.85$  \\
\hline
$\g_2$		&    $-0.92$			&  $0.84$		&   $-1.10$			&	$0.85$ \\
\hline
$\delta_1$	&   $23$				&  $21$		&   $21$				&	$23$ \\
\hline
$\delta_2$	&   $-200$				&  $48$		&    $-195$			&	$53$ \\
\hline
$\delta_3$	&   $278$				&  $30$		&    $268$				&	$34$ \\
\hline
$\delta_4$	&    $39$				&  $11$		&    $43$				&	$13$ \\
\hline
$\delta_5$	&    $-44$				&  $13$		&    $-48$				&	$13$ \\
\hline
$\eta$		&    $0.29$			&  $0.12$		&    $0.30$			&	$0.09$  \\
\hline
\end{tabular}
\caption{Average value $\{ \langle p \rangle \}$ and variance $\{ \sig_p \}$ of each element of the vector of fitting parameters 
$\{p \}$ in Eq.~(\ref{e:Hfit}), calculated by fitting $N=100$ replicas of the experimental data points for the Artru--Collins asymmetry (see discussion in the text around Eq.~(\ref{e:chi2A})). Left columns: QCD evolution performed with  
$\alpha_s (M_Z^2)=0.125$; right columns with  $\alpha_s(M_Z^2)=0.139$.} 
\label{tab:gaussparam} 
\end{center}
\end{table}

Finally, in Tab.~\ref{tab:gaussparam} we show the average value $\{ \langle p \rangle \}$ and variance $\{ \sig_p \}$ for each of the 9 elements in the vector $\{p \}$ of the fitting parameters in Eq.~(\ref{e:Hfit}). They are calculated from the set of $N$ different values obtained by fitting the $N$ replicas of the experimental data points, {\it i.e.} by minimizing the function $E_r^2(\{p\})$ in Eq.~(\ref{e:chi2A}) for $r=1, \ldots N$. The left pair of columns of numbers are obtained using $\alpha_s (M_Z^2)=0.125$, the right ones using $\alpha_s (M_Z^2)=0.139$. For some parameters, the variance is large compared to their average value, indicating that they are loosely constrained by the fit although the resulting $\chi^2$/d.o.f. are reasonable (see Fig.~\ref{fig:chi2}).


\section{Extraction of transversity}
\label{s:h1}

The valence components of transversity are extracted by combining Eqs.~(\ref{e:h1p}) and~(\ref{e:h1D}). In these equations, there are three main external ingredients: the unpolarized distributions $f_1^q$, the single-spin asymmetries $A^{p/D}_{{\rm SIDIS}}$, and the integrals $n_q$ and $n^\uparrow_u$. The unpolarized distributions $f_1^q$ are taken from the MSTW08 
set~\cite{Martin:2009iq} at leading order (LO). 

The experimental data for the single-spin asymmetries $A^p_{{\rm SIDIS}}$ and $A^D_{{\rm SIDIS}}$ are taken from the HERMES and COMPASS measurements on di-hadron SIDIS production off transversely polarized proton and deuteron targets. In our previous extraction~\cite{Bacchetta:2012ty}, we used the HERMES data for a proton target from 
Ref.~\cite{Airapetian:2008sk}, and the COMPASS data for unidentified hadron pairs $h^+ h^-$ produced off deuteron and proton targets from Ref.~\cite{Adolph:2012nw} (corresponding to the 2004 and 2007 runs, respectively). In an intermediate step~\cite{Radici:2014jra}, we have updated our extraction by using the new COMPASS data for unidentified hadron pairs $h^+ h^-$ produced off protons~\cite{Adolph:2014fjw}, corresponding to the 2010 run. In this work, we select the new COMPASS data for identified $\pi^+ \pi^-$ pairs produced off proton targets~\cite{Braun:2015baa}, still corresponding to the 2010 run. More precisely, we have used the results of Tab.~A.10 and~A.26 presented in the appendix~A.4 of 
Ref.~\cite{Braun:2014} where the proton data of the 2010 run have been complemented with a homogeneous re-analysis of all identified pairs from the 2004 and 2007 runs using the same data selection, methods, and binning. 

Finally, the third ingredient is represented by the integrals $n_q$ and $n^\uparrow_u$ of Eqs.~(\ref{e:nq}) 
and~(\ref{e:nqperp}), respectively. They are evaluated according to the appropriate experimental cuts: $0.2<z<1$ and $0.5\text{ GeV}<M_h<1$ GeV for HERMES, $0.2<z<1$ and $0.29\text{ GeV}<M_h<1.29$ GeV for COMPASS. The arguments of the integrals, namely the DiFFs $H_1^{\sphericalangle\, u}$ and $D_1^q$ with $q=u,d,s,c$, are determined along the lines described in the previous section. 

Equations~(\ref{e:h1p}) and~(\ref{e:h1D}) are fitted using the same strategy adopted in our previous extraction of 
Ref.~\cite{Bacchetta:2012ty}, and here used also for the extraction of DiFFs (see previous section). Namely, for each bin 
$(x_i,Q_i^2)$ a set of $M$ replicas of the data points $A^p_{{\rm SIDIS}}$ and $A^D_{{\rm SIDIS}}$ is created introducing a Gaussian noise with the same variance as the corresponding experimental measurement. Using the above information for the other ingredients $f_1^q$, $n_q$ and $n^\uparrow_u$, we build $M$ different replicas of Eqs.(\ref{e:h1p}) 
and~(\ref{e:h1D}) that we indicate with $x_i\, h_{1,r\, \text{exp}}^p (x_i, Q_i^2)$ and $x_i\, h_{1,r\, \text{exp}}^D (x_i, Q_i^2)$, respectively, with $r=1,\dots M$. Again, we checked that with $M=N=100$ replicas the mean and standard deviation of the original data points are accurately reproduced. In principle, for each bin $(x_i,Q_i^2)$ we can arbitrarily select one out of the $N$ possible values of  $n_q (Q_i^2)$ and $n^\uparrow_u (Q_i^2)$ obtained from the $N$ different replicas of DiFFs at that scale. So, in order to build $x_i\, h_{1,r\, \text{exp}}^{p/D} (x_i, Q_i^2)$ we simply pick up the replica $r$ of DiFFs, we calculate the corresponding integrals at the scale $Q_i^2$, and we associate them to the replica $r$ of the data points 
$A^{p/D}_{{\rm SIDIS}}$ in the bin $(x_i,Q_i^2)$. Then, for each replica $r$ we separately minimize an error function similar to Eq.~(\ref{e:chi2A}): 
\begin{equation}
E_r^{\prime \,2}(\{p' \}) = \sum_{i}  \frac{\left[ x_i\, h_{1, \text{th}}^{p/D} (x_i, Q_i^2, \{p' \})-  x_i\, h_{1,r\, \text{exp}}^{p/D} (x_i, Q_i^2)\right]^2}{\Big( \Delta h_{1\,\text{data}}^{p/D} (x_i, Q_i^2) \Big)^2 } \; , 
\label{e:chi2xh}
\end{equation}
where $\Delta h_{1\,\text{data}}^{p/D} (x_i, Q_i^2)$ are the errors on the original data points $A^{p/D}_{{\rm SIDIS}}$ multiplied by all factors according to Eqs.~(\ref{e:h1p}) and~(\ref{e:h1D}), in close analogy with $\sig_{ij}$ in 
Eq.~(\ref{e:chi2A}). 

The function $x_i\, h_{1, \text{th}}^{p/D} (x_i, Q_i^2, \{p' \})$ in Eq.~(\ref{e:chi2xh}) is obtained by evolving at the scale $Q_i^2$ of each bin a fitting function that depends on the vector of parameters $\{p' \}$. The main theoretical constraint that transversity must satisfy at each scale is Soffer's inequality~\cite{Soffer:1995ww}. If it is verified at some initial $Q_0^2$, it will do also at higher $Q^2 \geq Q_0^2$~\cite{Vogelsang:1997ak}. Therefore, following our previous work~\cite{Bacchetta:2012ty} we have parametrized the fitting function for the valence flavors $u_v$ and $d_v$ at $Q_0^2 = 1$ GeV$^2$ as
\beq
x\, h_1^{q_v}(x; Q_0^2) &= &\tanh \Bigl[ x^{1/2} \, \bigl( A_q+B_q\, x+ C_q\, x^2 +D_q\, x^3\bigr)\Bigr]\, x \, 
\Bigl[ \mbox{\small SB}^q(x; Q_0^2)+ \mbox{\small SB}^{\bar q}(x; Q_0^2)\Bigr] \, , \nn \\
& &
\label{e:h1fit}
\eeq
where the analytic expression of the Soffer bound SB$^q(x; Q^2)$ can be found in the Appendix of Ref.~\cite{Bacchetta:2012ty}. The implementation of the Soffer bound depends on the choice of the unpolarized PDF, as mentioned above, and of the helicity PDF that we take from the DSSV parameterization~\cite{deFlorian:2009vb}. As in Ref.~\cite{Bacchetta:2012ty}, the error on the Soffer bound is dominated by the uncertainty on the helicity $g_1$, that we checked to be negligible with respect to the experimental errors on $A_{\mathrm{SIDIS}}^{p/D}$. In Eq.~(\ref{e:h1fit}), the hyperbolic tangent is such that the Soffer bound is always fulfilled. The low-$x$ behavior is determined by the $x^{1/2}$ term, which is imposed by hand to grant that the resulting tensor charge is finite. Present fixed-target data do not allow to constrain it. This choice has little effect on the region where data exist, but has a crucial influence on extrapolations at low $x$. The functional form in Eq.~(\ref{e:h1fit}) is very flexible and can display up to three possible nodes. In analogy with Ref.~\cite{Bacchetta:2012ty}, we have explored three different scenarios: a) the {\it rigid} scenario, where $C_q = D_q = 0$ and the vector of parameters $\{p' \}$ contains only 4 free parameters; b) the {\it flexible} scenario, with $D_q = 0$ and 6 free parameters; c) the {\it extraflexible} scenario, with all possible 8 free parameters. The function $x_i\, h_{1, \text{th}}^{p/D} (x_i, Q_i^2, \{p' \})$ in Eq.~(\ref{e:chi2xh}) is then built by taking the proper flavor combinations of the fitting function in Eq.~(\ref{e:h1fit}) and evolving them to the scale $Q_i^2$ of each bin $i$. Evolution is realized using the {\tt HOPPET} code, suitably extended to include chiral-odd splitting functions at leading order~\cite{Courtoy:2012ry}, and with both values of $\alpha_s (M_Z^2)=0.125$ and $\alpha_s (M_Z^2)=0.139$.

\begin{figure}
\centering
\includegraphics[width=7cm]{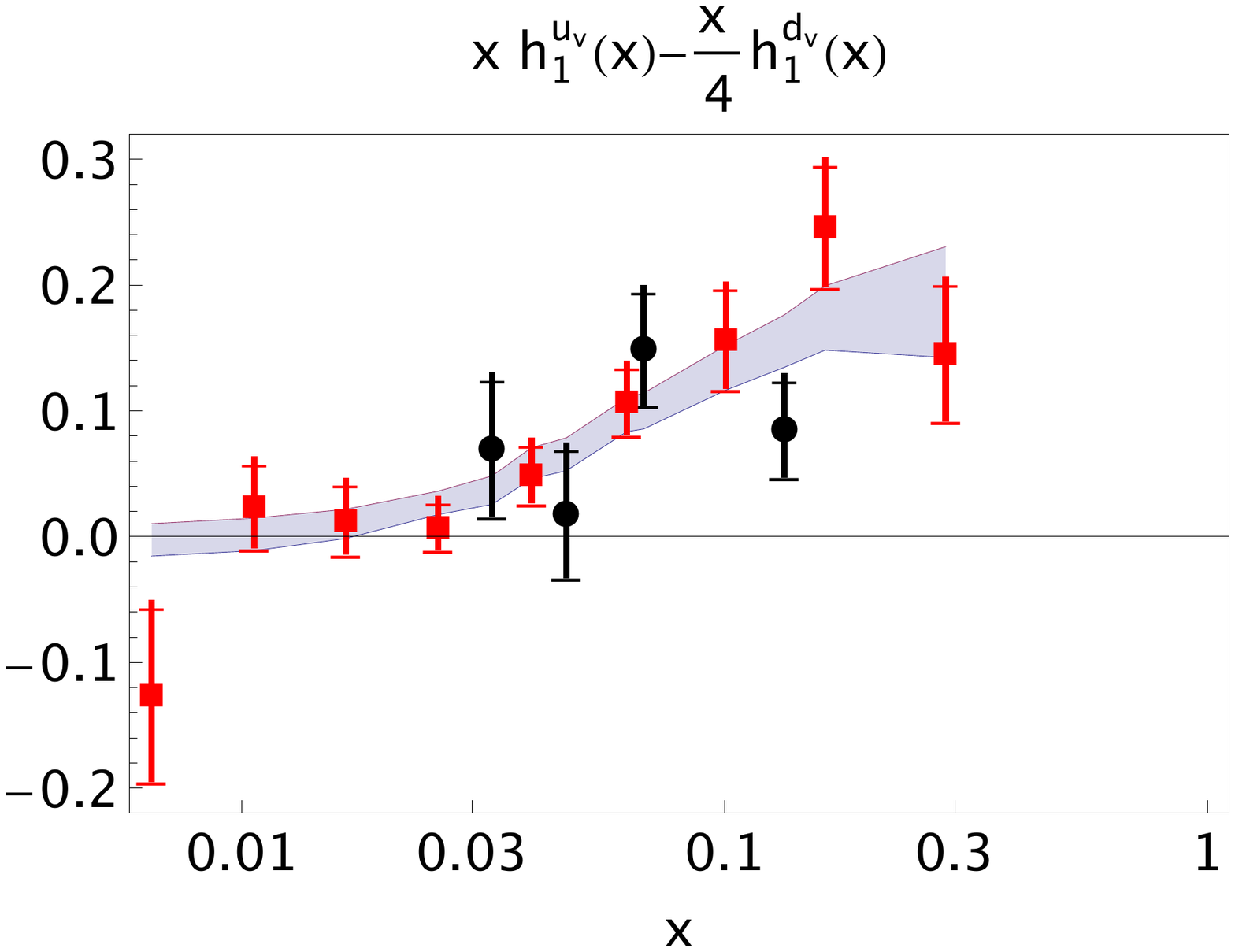} \hspace{0.5cm} \includegraphics[width=7cm]{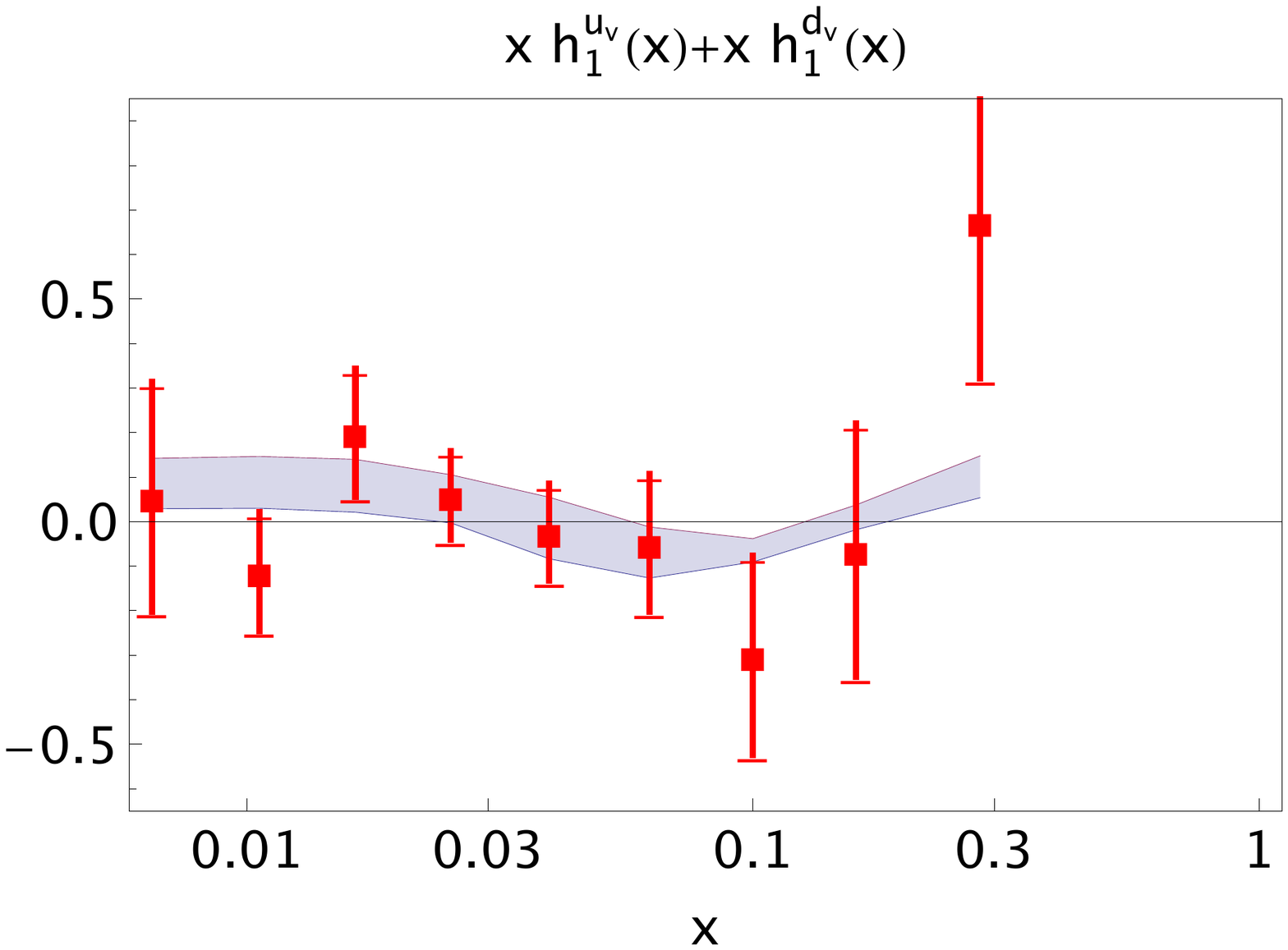}
\caption{The combinations of Eq.~(\ref{e:h1p}), left panel, and Eq.~(\ref{e:h1D}), right panel. The black circles are obtained from the HERMES data for the SSA $A^p_{{\rm SIDIS}}$; the lighter squares from the COMPASS data for both 
$A^p_{{\rm SIDIS}}$ and $A^D_{{\rm SIDIS}}$. The uncertainty band represents the selected $68\%$ of all fitting replicas in the rigid scenario with $\alpha_s (M_Z^2)=0.125$ (see text).}
\label{fig:xh1pDfit}
\end{figure}

In Fig.~\ref{fig:xh1pDfit}, the points represent the combinations of Eq.~(\ref{e:h1p}) in the left panel, and of Eq.~(\ref{e:h1D}) in the right panel. The error bars are mainly determined by the experimental errors on $A^p_{{\rm SIDIS}}$ and 
$A^D_{{\rm SIDIS}}$, respectively, because the uncertainty on the extracted DiFFs is much smaller. The black circles in the left panel are obtained when using for $A^p_{{\rm SIDIS}}$ the HERMES measurement from Ref.~\cite{Airapetian:2008sk}. The lighter squares in both panels correspond to the COMPASS measurements of $A^p_{{\rm SIDIS}}$ (left) and 
$A^D_{{\rm SIDIS}}$ (right) from Ref.~\cite{Braun:2014}, as explained above. The uncertainty bands show the result of the $68\%$ of all fitting replicas in the rigid scenario with $\alpha_s (M_Z^2)=0.125$. They are obtained by minimizing the error function in Eq.~(\ref{e:chi2xh}) and by further rejecting the largest $16\%$ and the lowest $16\%$ of the $M=100$ replicas' values in each $x$ point.

\begin{figure}
\centering
\includegraphics[width=7cm]{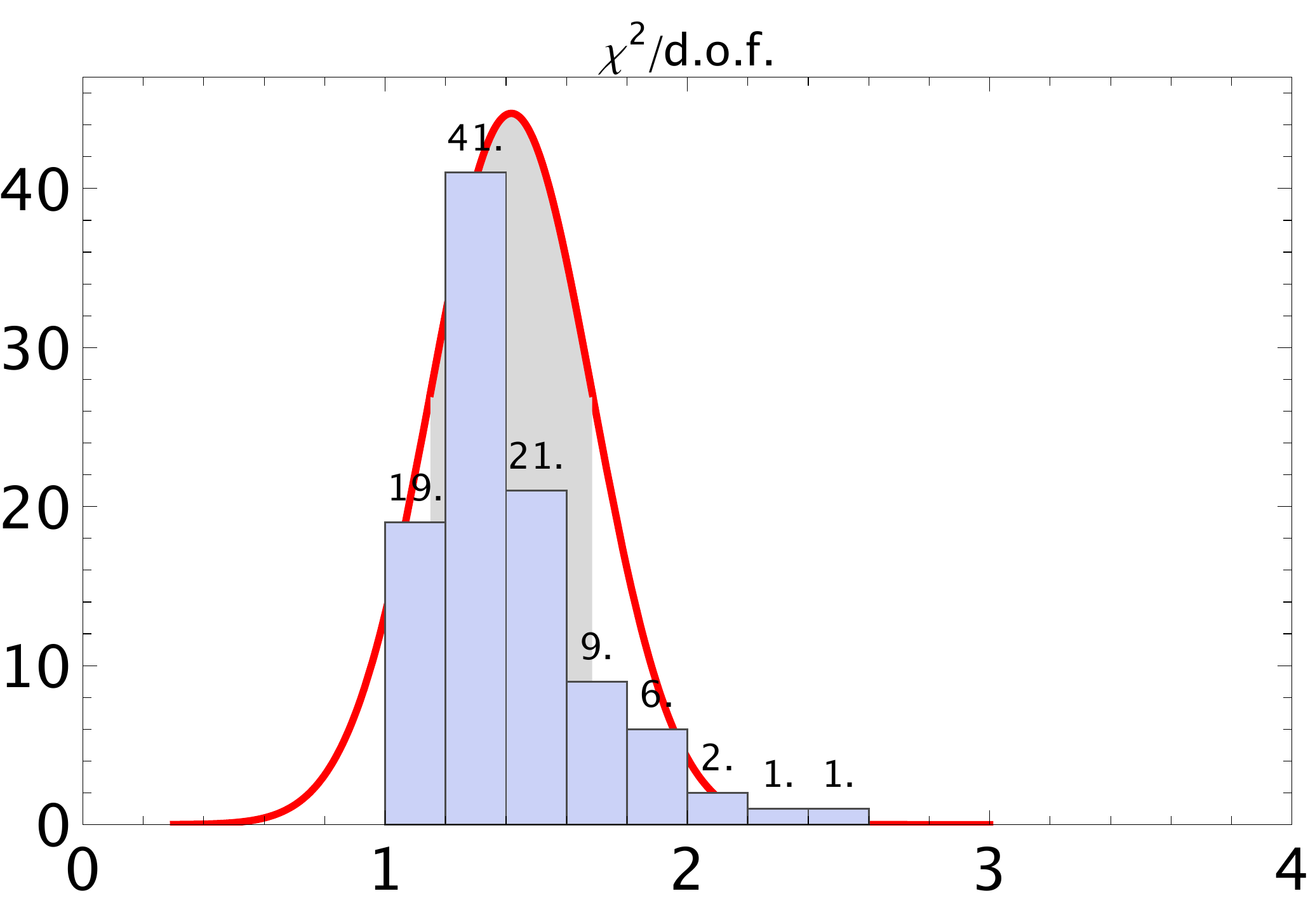}
\caption{Histogram of the distribution of $M=100$ $\chi^2$/d.o.f. when minimizing Eq.~(\ref{e:chi2xh}) for 
$\alpha_s (M_Z^2)=0.125$ and in the rigid scenario. The solid curve corresponds to a Gaussian distribution centered at the average of the $M=100$ $\chi^2$ values. The shaded area represents the $1\sig$ variance. The normalization of the Gaussian distribution is adapted to the histogram profile.}
\label{fig:chi2rigid}
\end{figure}

In Fig.~\ref{fig:chi2rigid}, we show the histogram for the distribution of the $M$ values of the $\chi^2$/d.o.f obtained by minimizing the error function in Eq.~(\ref{e:chi2xh}) for the rigid scenario with $\alpha_s (M_Z^2)=0.125$. For sake of illustration, we compare it with the solid line representing a Gaussian distribution centered around the average 1.42 of the $\chi^2$/d.o.f. values for this scenario. The shaded area represents the $1\sig$ variance. The distribution is not peaked at 1 but around 1.4 because of the rigidity of the fitting model. When changing evolution parameter from $\alpha_s (M_Z^2)=0.125$ to $\alpha_s (M_Z^2)=0.139$, the salient features of the $\chi^2$ distribution remain substantially the same and the average $\chi^2$/d.o.f. increases by less than 3\%, as it can be realized by inspecting Tab.~\ref{tab:chi2xhfit}.

\begin{table}[h]
\begin{center}
\begin{tabular}{ |c||c|c| }
  \hline
$\chi^2$/d.o.f.	&  $\alpha_s (M_Z^2)=0.125$ 	&  $\alpha_s (M_Z^2)=0.139$    \\
  \hline
  \hline	
rigid		&	1.42 					& 1.46 	   \\
\hline
flexible	&    1.65					& 1.71 	  \\
\hline
extraflexible &   1.97					&  2.07	 \\
\hline
\end{tabular}
\caption{The average $\chi^2$/d.o.f. obtained by minimizing the error function in Eq.~(\ref{e:chi2xh}) for the three different scenarios explored in the fitting function, and for the two values of $\alpha_s$ in the evolution code.} 
\label{tab:chi2xhfit} 
\end{center}
\end{table}

\begin{figure}
\centering
\includegraphics[width=8cm]{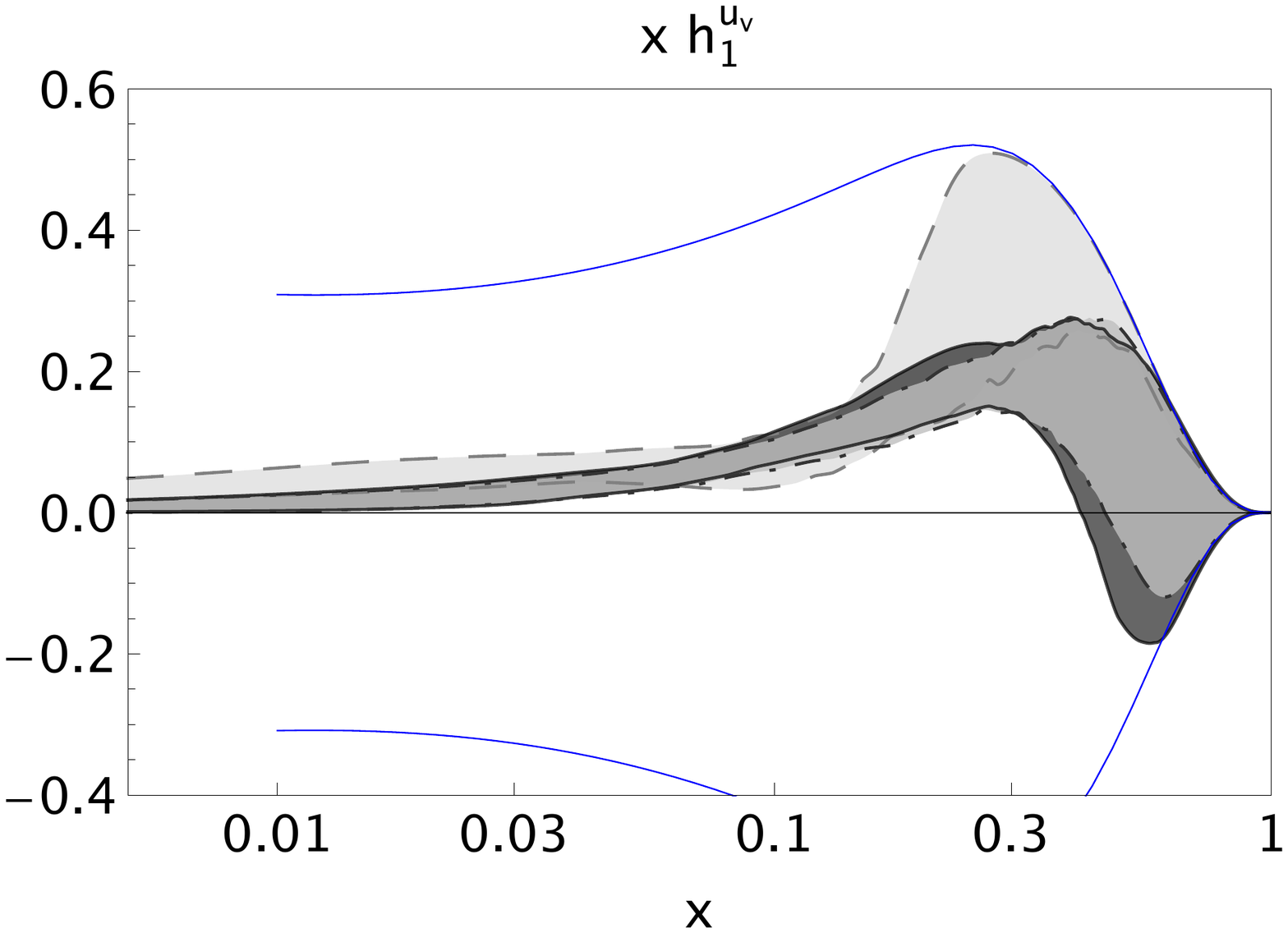}
\caption{The up valence transversity as a function of $x$ at $Q^2=2.4$ GeV$^2$ in the flexible scenario. The brightest band in the background with dashed borders is the 68\% of all replicas from our previous extraction~\cite{Bacchetta:2012ty}. The light grey band in the foreground with dot-dashed borders is the 68\% of all replicas obtained in this work with 
$\alpha_s (M_Z^2)=0.139$. The darkest band with solid borders is the same but for $\alpha_s (M_Z^2)=0.125$. The thick solid lines indicate the Soffer bound.}
\label{fig:xhuflex}
\end{figure}

In Fig.~\ref{fig:xhuflex}, we show the up valence transversity, $x h_1^{u_v}$, as a function of $x$ at $Q^2=2.4$ GeV$^2$ in the flexible scenario. The brightest band in the background with dashed borders is the 68\% of all replicas from our previous 
extraction~\cite{Bacchetta:2012ty}. The light grey band in the foreground with dot-dashed borders shows the 68\% of all replicas obtained in this work when using $\alpha_s (M_Z^2)=0.139$. The darkest band with solid borders is the result when using $\alpha_s (M_Z^2)=0.125$. Finally, the thick solid lines indicate the Soffer bound. The fact that the latter two bands overlap almost completely confirms that our new extraction is not very sensitive to the value of $\alpha_s (M_Z^2)$, namely to the theoretical uncertainty in the evolution equations. On the other side, the impact of the new COMPASS data is rather evident. There is still overlap between present and previous extractions, but the better statistical precision of data produces a narrower uncertainty band, at least in the range $0.0065 \leq x \leq 0.29$ where there are data. Moreover, the replicas spread out over values that on average are smaller than before. Since the new COMPASS analysis of Ref.~\cite{Adolph:2014fjw} deals with proton targets, the combination in Eq.~(\ref{e:h1D}) is not affected. Our extraction of the down valence transversity is basically unchanged with respect to the previous one~\cite{Bacchetta:2012ty}; therefore, we will not show it. Similar results are obtained when switching to other scenarios in the fitting function; we will not show them as well.

\begin{figure}
\centering
\includegraphics[width=7cm]{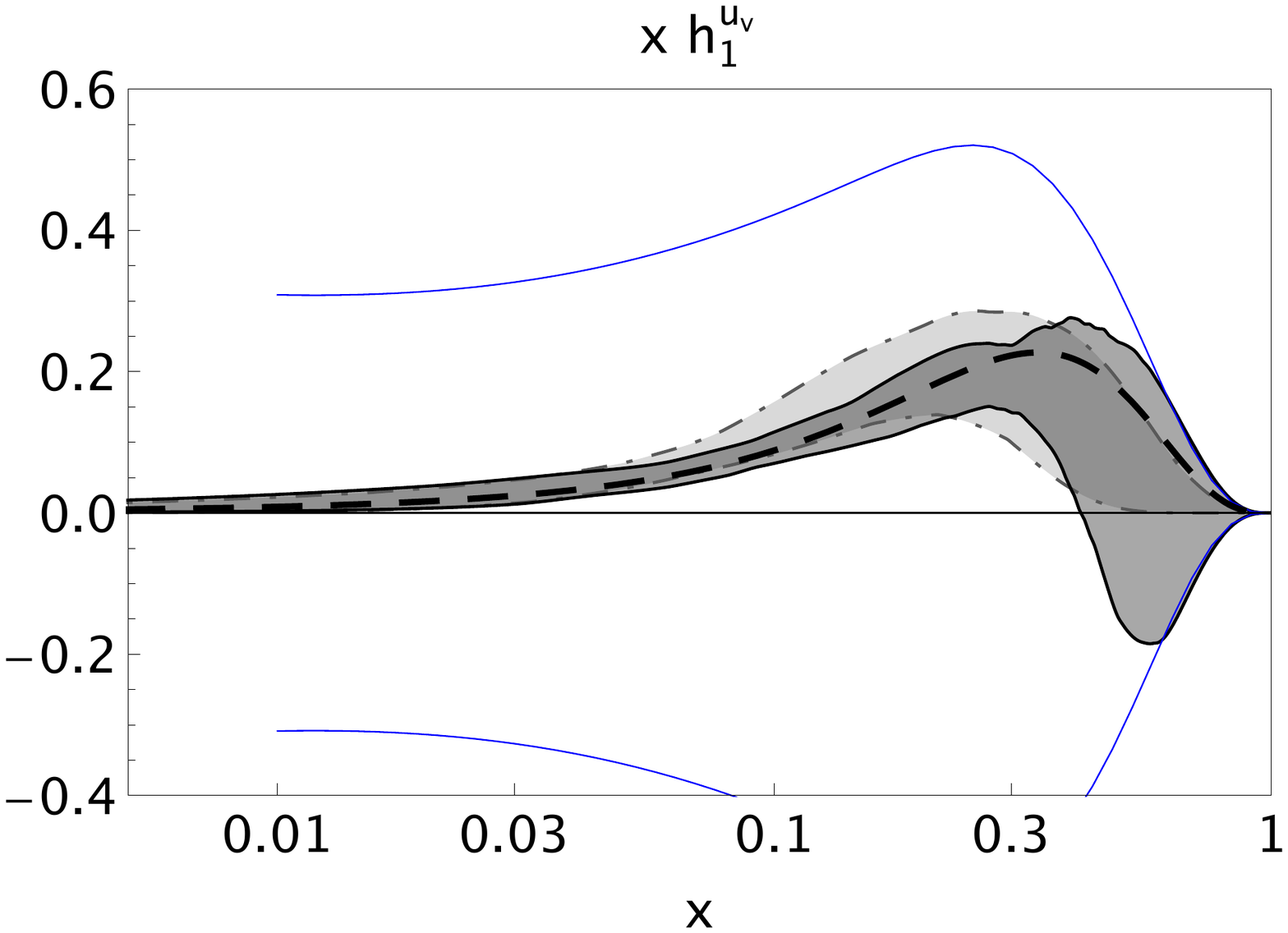} \hspace{0.5cm} \includegraphics[width=7cm]{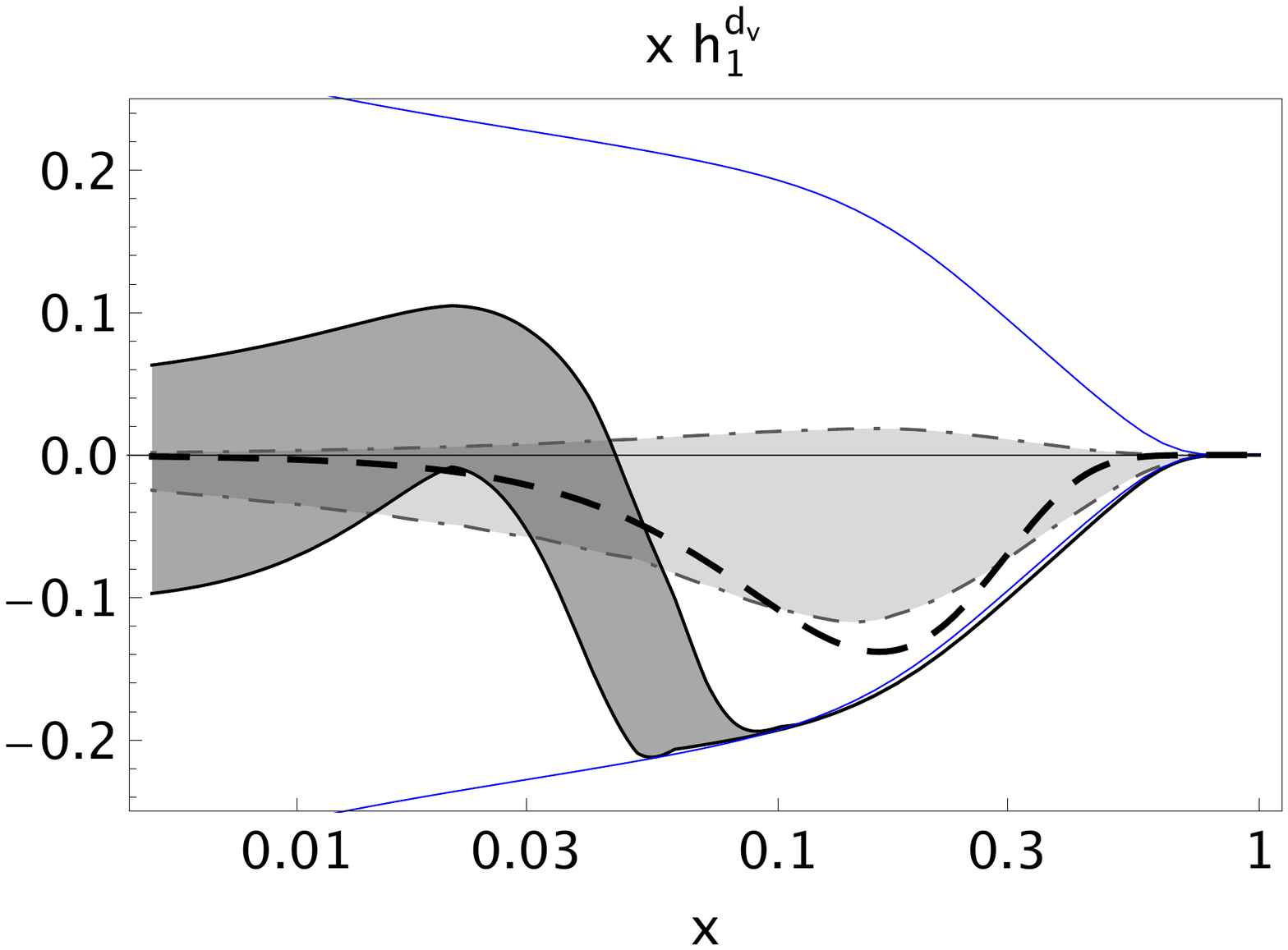}
\caption{The up (left) and down (right) valence transversities as functions of $x$ at $Q^2=2.4$ GeV$^2$. The darker band with solid borders in the foreground is our result in the flexible scenario with $\alpha_s (M_Z^2)=0.125$. The lighter band with dot-dashed borders in the background is the most recent transversity extraction from the Collins effect~\cite{Anselmino:2013vqa}. The central thick dashed line is the result of Ref.~\cite{Kang:2014zza}. The thick solid lines indicate the Soffer bound.}
\label{fig:xhud}
\end{figure}

In Fig.~\ref{fig:xhud}, we show how our new results compare with other extractions of transversity based on the Collins effect. In the left (right) panel, the up (down) valence transversity is displayed as a function of $x$ at $Q^2=2.4$ GeV$^2$. The darker band with solid borders in the foreground is our result in the flexible scenario with $\alpha_s (M_Z^2)=0.125$. The lighter band with dot-dashed borders in the background is the most recent transversity extraction of Ref.~\cite{Anselmino:2013vqa} using the Collins effect but applying the standard DGLAP evolution equations only to the collinear part of the fitting function. The central thick dashed line is the result of Ref.~\cite{Kang:2014zza}, where evolution equations have been computed in the TMD framework. 

In the right panel, the disagreement between our result for $x h_1^{d_v} (x)$ at $x \geq 0.1$ and the outcome of the Collins effect is confirmed  with respect to our previous analysis (see Fig.~4 in Ref.~\cite{Bacchetta:2012ty}). This is due to the fact that the COMPASS data for $A^D_{{\rm SIDIS}}$ off deuteron targets remain the same. This trend is confirmed also in the other scenarios, indicating that it is not an artifact of the chosen functional form. As a matter of fact,  our replicas for the valence down transversity tend to saturate the lower limit of the Soffer bound because they are driven by the COMPASS deuteron data, in particular by the bins number 7 and 8. It is worth mentioning that some of the replicas outside the 68\% band do not follow this trend. Their trajectories are spread over the whole available space between the upper and lower limits of the Soffer bound, still maintaining a good $\chi^2$/d.o.f. (typically, around 2). It is also interesting to remark that the dashed line from Ref.~\cite{Kang:2014zza}, although in general agreement with the other extraction based on the Collins effect, also tends to saturate the Soffer bound at $x > 0.2$. 

Apart from the range $x \geq 0.1$, there is a general consistency among the various extractions which is confirmed also for the valence up transversity (left panel), at least for the range $0.0065 \leq x \leq 0.29$ where there are data. This is encouraging: while the dihadron SIDIS data are a subset of the single-hadron ones, the theoretical frameworks used to interpret them are very different. Nevertheless, we point out that the collinear framework, in which our results are produced, represents a well established and robust theoretical context. On the contrary, the implementation of the QCD evolution equations of TMDs needed in the study of the Collins effect still contains elements of arbitrariness (see Refs.~\cite{Collins:2014loa,Echevarria:2014rua,Kang:2014zza} and references therein). Moreover, we believe that our error analysis, based on the replica method applied to the extraction of both the DiFFs from $e^+ e^-$ data and the transversity from SIDIS data, represents the current most realistic estimate of the uncertainties on transversity. It also clearly shows that we have no clue on the transversity for large 
$x \geq 0.3$ where there are no data at present. This is particularly evident in the left panel of Fig.~\ref{fig:xhud}: the replicas in the darker band tend to fill all the available phase space within the solid lines of the Soffer bound, graphically visualizing our poor knowledge of $x h_1^{u_v} (x)$ in that range. Similarly, data are missing also for very small $x$, and this prevents from fixing the behaviour of transversity for $x \to 0$ in a less arbitrary way than the choice made in Eq.~(\ref{e:h1fit}).

\begin{figure}
\centering
\includegraphics[width=7cm]{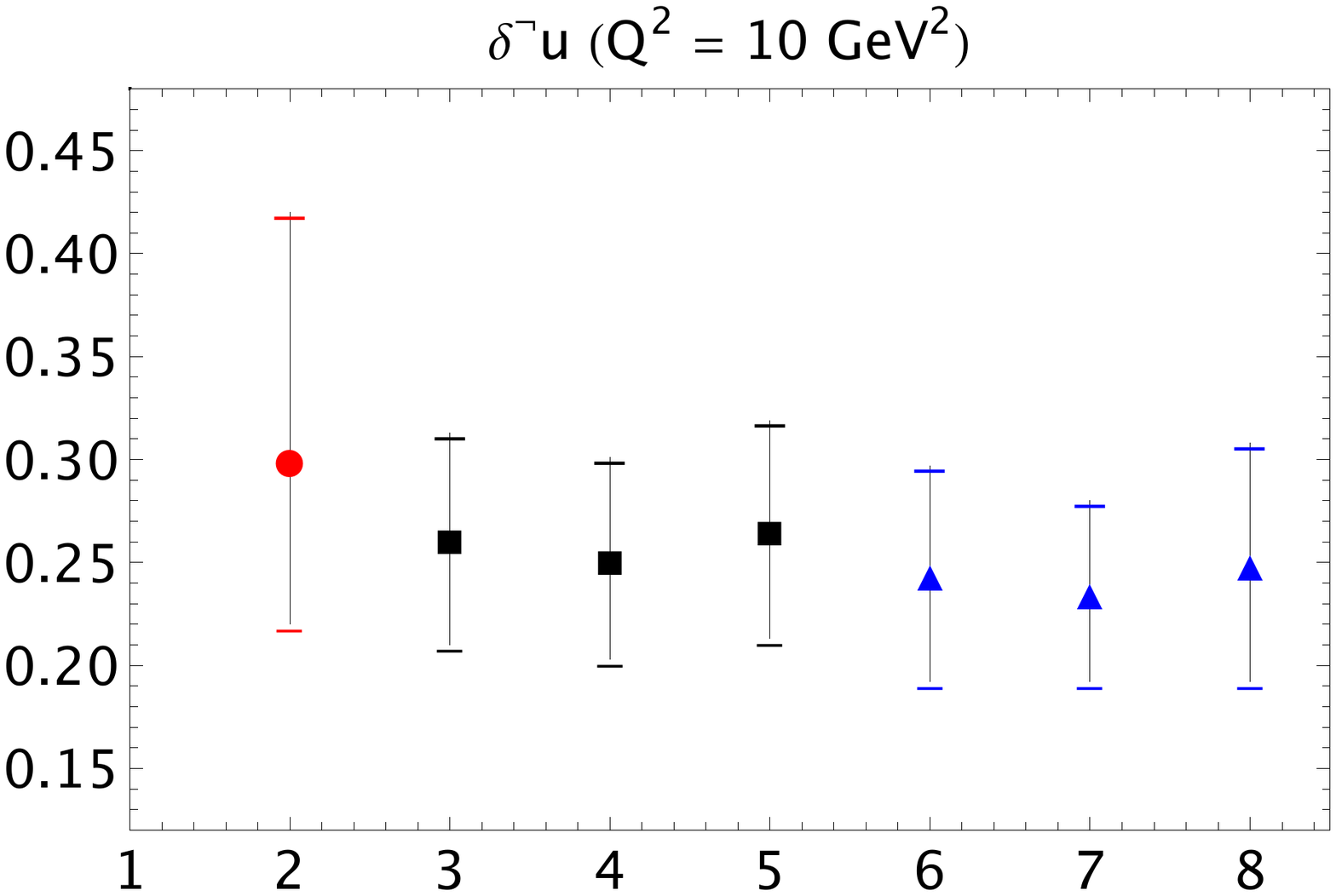} \hspace{0.5cm} \includegraphics[width=7cm]{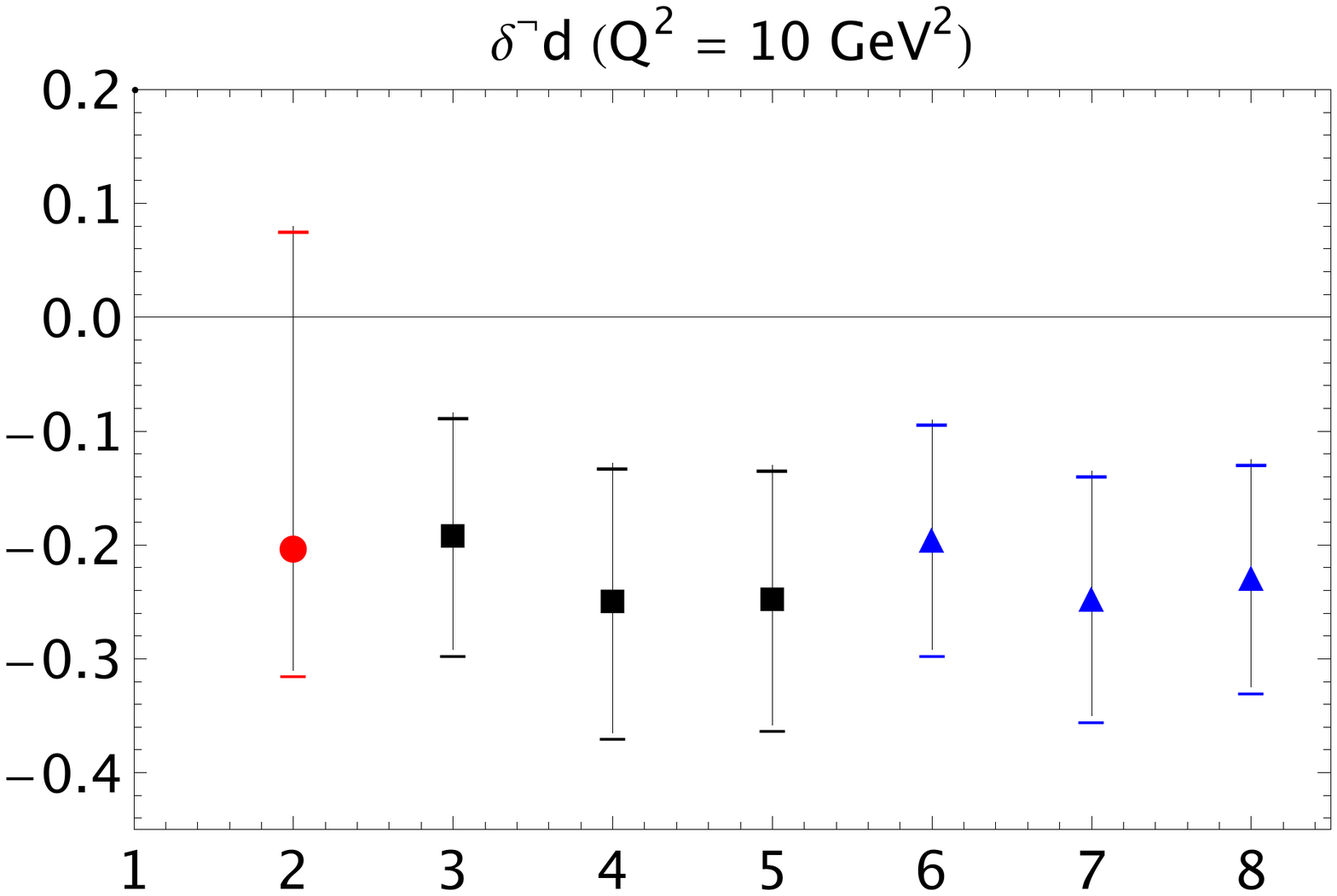}
\caption{Truncated tensor charges (see text) at $Q^2=10$ GeV$^2$ for the valence up (left panel) and down quark (right panel). From left to right: circle (label 2) for the value obtained through the Collins effect in Ref.~\cite{Kang:2014zza}, black squares (labels 3-5) for the rigid, flexible, extraflexible scenarios, respectively, here explored with $\alpha_s (M_Z^2)=0.125$, triangles (labels 6-8) for the corresponding ones with $\alpha_s (M_Z^2)=0.139$.}
\label{fig:tensor-data}
\end{figure}

In Fig.~\ref{fig:tensor-data}, we show the "truncated" tensor charge
\beq
\delta\urcorner q_v (Q^2) &= &\int_{x_{\text{min}}}^{x_{\text{max}}} dx \, h_1^{q_v} (x, Q^2) \; , 
\label{e:trunctensch}
\eeq
namely the truncated first Mellin moment of the valence transversity. The integral is computed for $x_{\text{min}} = 0.0065 \leq x \leq x_{\text{max}} = 0.29$, {\it i.e.} in the range of experimental data, thus avoiding any numerical uncertainty produced by extrapolation outside this range. In the left panel, we show $\delta\urcorner u_v (Q^2=10$ GeV$^2)$, in the right panel $\delta\urcorner d_v (Q^2=10$ GeV$^2)$. They are calculated at $Q^2=10$ GeV$^2$ in order to compare with the results of Ref.~\cite{Kang:2014zza}, which are indicated in both panels by the leftmost circle with label 2. The black squares with labels 3-5 indicate our result with $\alpha_s (M_Z^2)=0.125$ for the rigid, flexible, and extraflexible scenarios, from left to right respectively. The triangles with labels 6-8 correspond to the choice $\alpha_s (M_Z^2)=0.139$ in the same order. The corresponding error bars are computed by considering the distance between the minimum and the maximum values of the 68\% of all replicas; the squares and triangles identify their equidistant point. Our results are basically insensitive to the choice of $\alpha_s$; so, in the following we will show results only for the choice $\alpha_s (M_Z^2)=0.125$, forwarding the reader to Tab.~\ref{tab:all} for the numerical values of all considered cases.

\begin{figure}
\centering
\includegraphics[width=7cm]{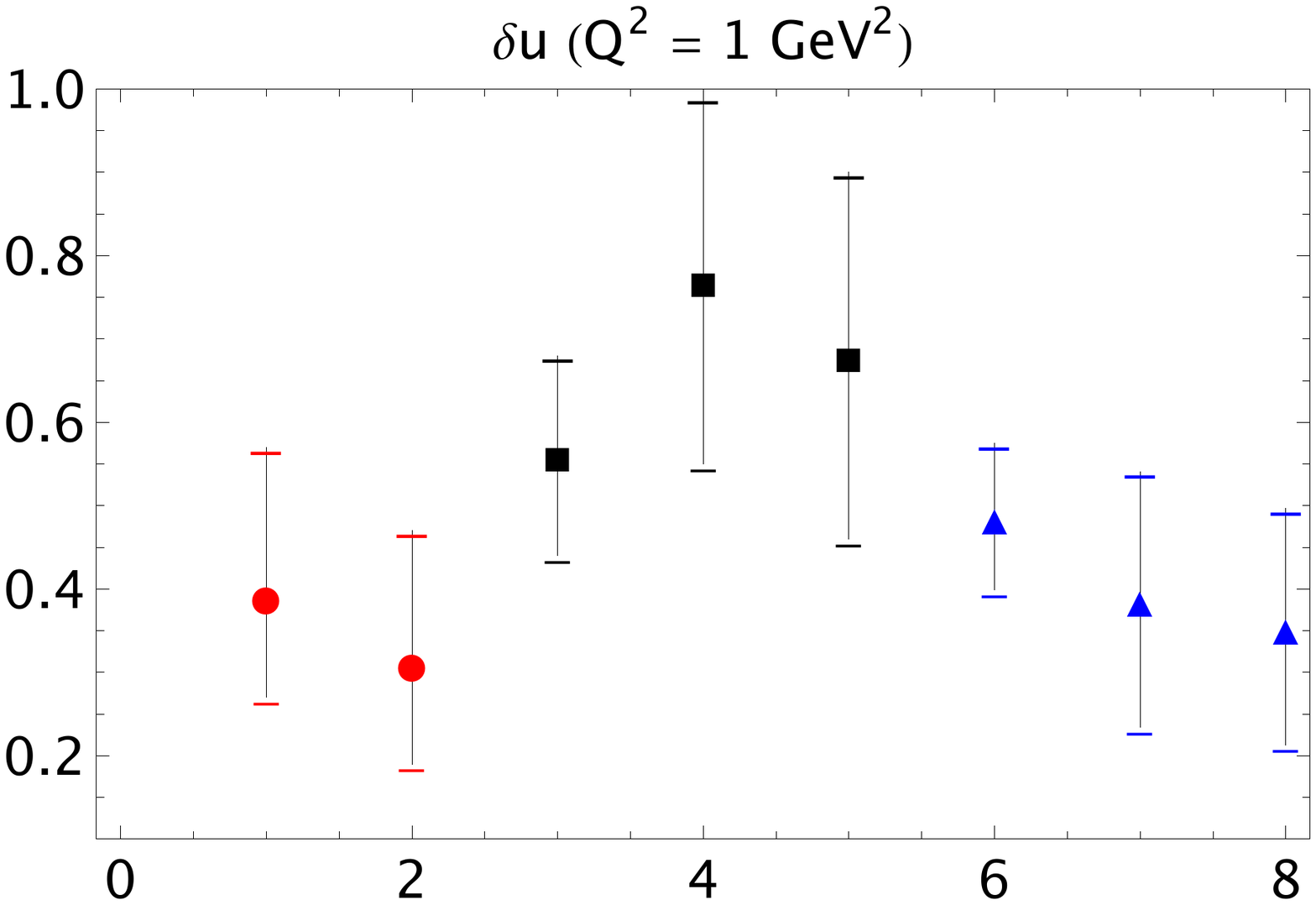} \hspace{0.5cm} \includegraphics[width=7cm]{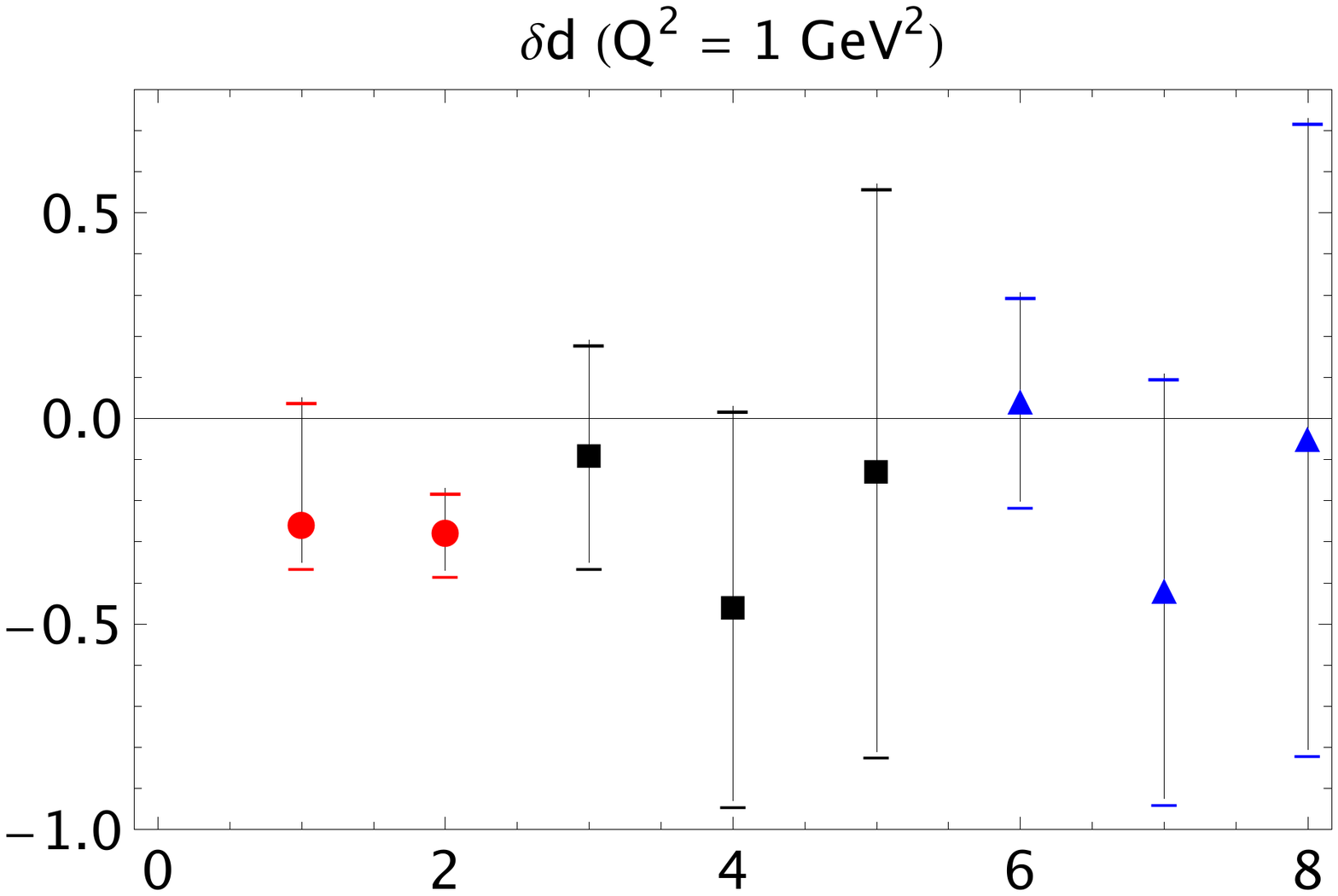}
\caption{Tensor charges at $Q_0^2=1$ GeV$^2$ for the valence up (left panel) and down quark (right panel). From left to right: circles (label 1, 2) for the values obtained through the Collins effect in Ref.~\cite{Anselmino:2013vqa}, black squares (labels 3-5) for the rigid, flexible, extraflexible scenarios explored in our previous extraction of Ref.~\cite{Bacchetta:2012ty}, triangles (labels 6-8) for the present work with $\alpha_s (M_Z^2)=0.125$.}
\label{fig:tensor-full}
\end{figure}

In Fig.~\ref{fig:tensor-full}, we show the full Mellin moments of valence transversity at $Q_0^2=1$ GeV$^2$, {\it i.e.} the tensor charges
\beq
\delta q_v (Q^2) &= &\int_0^1 dx \, h_1^{q_v} (x, Q^2) \; . 
\label{e:tensch}
\eeq
The integration is now extended to the full $x$ domain by extrapolating $h_1^{q_v} (x)$ outside the experimental range. As in the previous figure, the left panel refers to the valence up quark while the right one to the valence down quark. The two leftmost circles (labels 1, 2) are the results obtained from the analysis of the Collins effect using two different methods for the extraction of the Collins function from $e^+ e^-$ annihilation data~\cite{Anselmino:2013vqa}. The three black squares (labels 3-5) correspond to the results of our previous analysis~\cite{Bacchetta:2012ty} for the rigid, flexible, and extraflexible scenarios, from left to right respectively. The three rightmost triangles (labels 6-8) indicate the outcome of the present work with 
$\alpha_s (M_Z^2)=0.125$ in the same order. Consistently with Fig.~\ref{fig:xhuflex}, our new results for the up quark are smaller than the previous ones. They also appear globally in better agreement with the values from Ref.~\cite{Anselmino:2013vqa} (and not far from the ones obtained from the parametrization of chiral-odd Generalized Parton Distributions of 
Ref.~\cite{Goldstein:2014aja}), although the large uncertainties introduced by the numerical extrapolation smooth most of the differences. This is particularly evident for the down quark, where in addition the numerical values are very close because the experimental data for $A^D_{{\rm SIDIS}}$ are the same as before. 

\begin{figure}
\centering
\includegraphics[width=8cm]{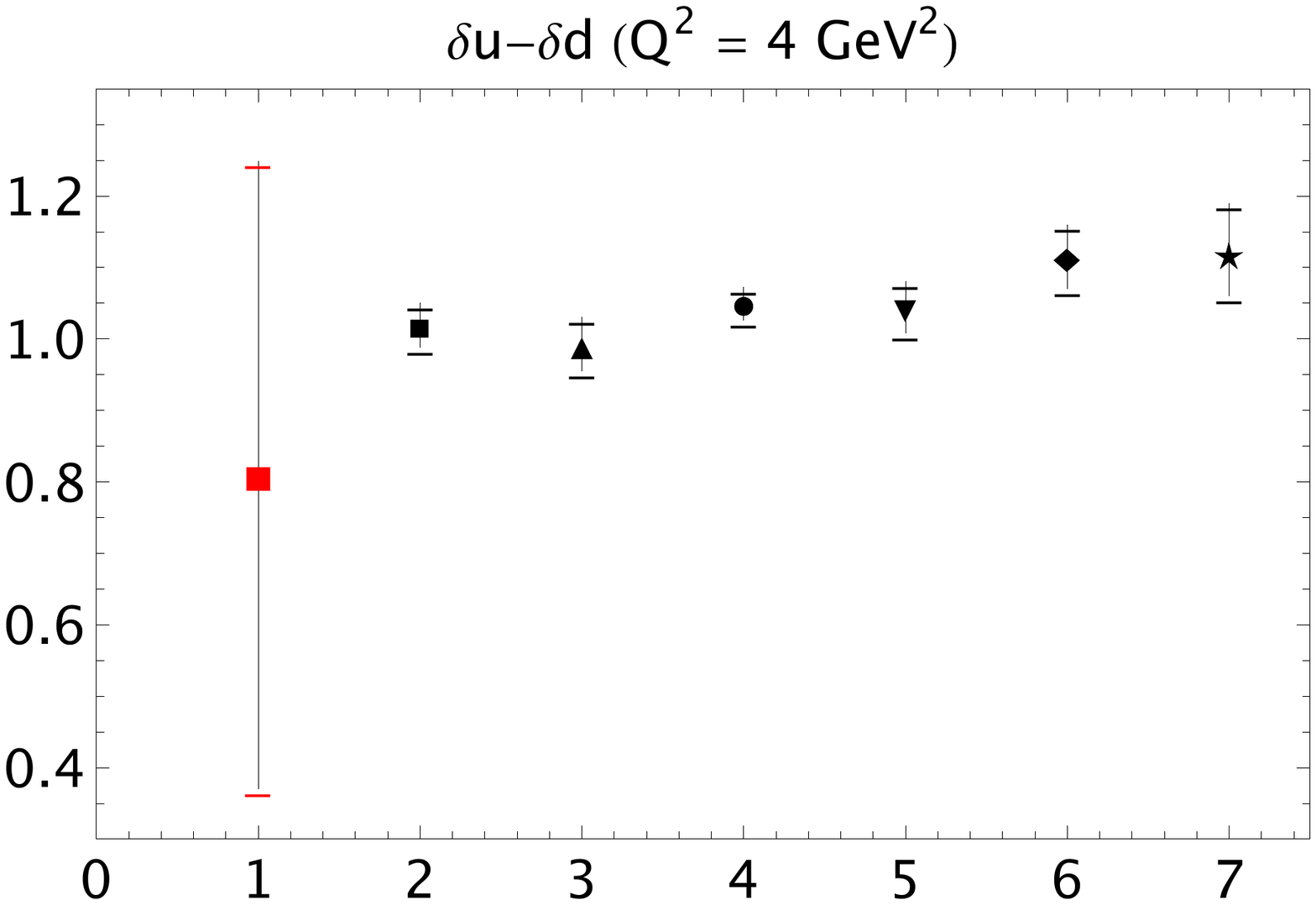} 
\caption{Isovector tensor charge $\delta u_v - \delta d_v$ at $Q^2=4$ GeV$^2$. From left to right: light square (label 1) is our result for the flexible scenario with $\alpha_s (M_Z^2)=0.125$; black square for the lattice result of Ref.~\cite{Bali:2014nma} (RQCD); black triangle from Ref.~\cite{Aoki:2010xg} (RBC-UKQCD); black circle from Ref.~\cite{Green:2012ej} (LHPC); black inverted triangle from Ref.~\cite{Bhattacharya:2013ehc} (PNDME); black diamond and star from Ref.~\cite{Alexandrou:2013wka} (ETMC) with 2+1 and 2+1+1 flavors, respectively.}
\label{fig:isotensor}
\end{figure}

In Fig.~\ref{fig:isotensor}, we show the isovector nucleon tensor charge $g_T = \delta u_v - \delta d_v$. While there is no elementary tensor current at tree level in the Standard Model, the nucleon matrix element of the tensor operator can still be defined (for a review, see Ref.~\cite{Barone:2003fy} and references therein). The $g_T$ belongs to the group of isovector nucleon charges that are related to flavour-changing processes. A determination of 
these couplings may shed light on the search of new physics mechanisms that may depend on 
them~\cite{Bhattacharya:2011qm,Ivanov:2012qe,Cirigliano:2013xha,Courtoy:2014xea}, or on direct dark matter 
searches~\cite{DelNobile:2013sia}. The vector charge $g_V$, axial charge $g_A$, and induced tensor charge $\tilde{g}_T$, are fixed by baryon number conservation, neutron $\beta$-decay, and nucleon magnetic moments, 
respectively~\cite{Agashe:2014kda}. Also the pseudoscalar charge $g_P$ is, to some extext, constrained by low-energy 
$n \pi^+$ scattering~\cite{Gasser:1987rb}. The other isovector nucleon couplings, including $g_T$, have been determined so far only with lattice QCD. 

In Fig.~\ref{fig:isotensor}, the leftmost light square with label 1 is our new result for $g_T = 0.81 \pm 0.44$ at $Q^2=4$ GeV$^2$ for the flexible scenario with $\alpha_s (M_Z^2)=0.125$. We compare it with various lattice computations. From left to right, the black square refers to the lattice simulation of RQCD at $m_\pi \approx 150$ MeV with $n_f = 2$ NPI Wilson-clover 
fermions~\cite{Bali:2014nma}, the black triangle to that of RBC-UKQCD at $m_\pi = 330$ MeV with $n_f = 2+1$ domain wall fermions~\cite{Aoki:2010xg}, the black circle to that of LHPC at $m_\pi \approx 149$ MeV with $n_f = 2+1$ HEX-smeared Wilson-clover fermions~\cite{Green:2012ej}, the black inverted triangle to that of PNDME at $m_\pi = 220$ MeV with Wilson-clover fermions on a HISQ staggered $n_f = 2+1+1$ sea~\cite{Bhattacharya:2013ehc}, the black diamond and star to that of ETMC at physical $m_\pi$ with $n_f = 2$ twisted mass fermions and at $m_\pi = 213$ MeV with $n_f = 2+1+1$ twisted mass fermions, respectively~\cite{Alexandrou:2013wka}. Our result is obviously compatible with the various lattice simulations because of the very large error. As already remarked, this originates from the fact that the integral in Eq.~(\ref{e:tensch}) involves the extrapolation of transversity outside the $x$ range of experimental data. From 
Fig.~\ref{fig:xhuflex} it is evident that the replicas tend to take all values within the Soffer bounds for $x\geq 0.3$ where there are no data, thus increasing the uncertainty. Moreover, we stress again that there is also a source of systematic error related to the power $x^{1/2}$ in the fitting form of Eq.~(\ref{e:h1fit}). The absence of data at very low $x$ leaves this choice basically unconstrained, whereas the value of the integral in Eq.~(\ref{e:tensch}) heavily depends on it.

\begin{table}[h]
\begin{tabular}{ |c||c|c||c|c| }
\hline
\multicolumn{1}{|c||}{\small{$\delta^{\urcorner} q_v (Q^2 = 10$ GeV$^2)$}}& \multicolumn{2}{|c||}{\small{valence up}}& \multicolumn{2}{|c|}{\small{valence down}} 
\\
\hline
 	              &\small{$\alpha_s(M_Z^2)=0.125$} &\small{$\alpha_s(M_Z^2)=0.139$} &\small{$\alpha_s(M_Z^2)=0.125$} &\small{$\alpha_s(M_Z^2)=0.139$}
\\
\hline
\hline	
\small{rigid}  &\small{$0.26 \pm 0.05$} 	&\small{$0.24 \pm 0.05$} &\small{$-0.19 \pm 0.10$}	 &\small{$-0.19 \pm 0.10$}  \\
\hline
\small{flexible}	&\small{$0.25 \pm 0.05$}&\small{$0.24 \pm 0.04$}&\small{$-0.25 \pm 0.12$} &\small{$-0.24 \pm 0.11$}  \\
\hline
\small{extraflexible}  &\small{$0.27 \pm 0.05$}&\small{$0.25 \pm 0.06$}&\small{$-0.24 \pm 0.11$} &\small{$-0.22 \pm 0.10$} \\
\hline
\hline
\multicolumn{5}{|c|}{} \\
\hline
\hline
\multicolumn{1}{|c||}{\small{$\delta q_v (Q_0^2 = 1$ GeV$^2)$}}     &      &          \\
\hline
\hline
\small{rigid}&\small{$0.49 \pm 0.09$} &\small{$0.43 \pm 0.08$}  &\small{$0.05 \pm 0.25$}&\small{$0.04 \pm 0.24$}  \\
\hline
\small{flexible}&\small{$0.39 \pm 0.15$}&\small{$0.40 \pm 0.14$}&\small{$-0.41 \pm 0.52$} &\small{$-0.32 \pm 0.51$}  \\
\hline
\small{extraflexible}  &\small{$0.35 \pm 0.14$}&\small{$0.36 \pm 0.12$}&\small{$-0.04 \pm 0.77$}&\small{$-0.12 \pm 0.74$} \\
\hline
\end{tabular}
\caption{Summary of numerical values for the tensor charge. Upper part for the truncated tensor charge of Eq.~(\ref{e:trunctensch}) at $Q^2=10$ GeV$^2$ for valence up and down quarks in the rigid, flexible, extraflexible scenarios for the fitting function of Eq.~(\ref{e:h1fit}) with $\alpha_s(M_Z^2)=0.125$ or $\alpha_s(M_Z^2)=0.139$ in the evolution code. Lower part for the tensor charge of Eq.~(\ref{e:tensch}) at $Q_0^2=1$ GeV$^2$.}
\label{tab:all} 
\end{table}

Finally, in Tab.~\ref{tab:all} we collect all numerical values that we have obtained for the (truncated) tensor charge. In the upper part of the table, we show the truncated tensor charge $\delta^{\urcorner} q_v$ of Eq.~(\ref{e:trunctensch}) at $Q^2=10$ 
GeV$^2$ for valence up and down quarks in the rigid, flexible, extraflexible scenarios for the fitting function of Eq.~(\ref{e:h1fit}) with $\alpha_s(M_Z^2)=0.125$ or $\alpha_s(M_Z^2)=0.139$ in the evolution code. In the lower part of the table, we show the results for the same cases but for the tensor charge $\delta q_v$ of Eq.~(\ref{e:tensch}) at the starting scale $Q_0^2=1$ 
GeV$^2$.
 

\section{Conclusions}
\label{s:end}

The transversity parton distribution function is an essential piece of information on the nucleon at leading twist. Its first Mellin moment is related to the nucleon tensor charge. Due to its chiral-odd nature, transversity cannot be accessed in fully inclusive deep-inelastic scattering (DIS). Within the framework of collinear factorization, it is however possible to access it in two-particle-inclusive DIS in combination with Dihadron Fragmentation Functions (DiFFs). The latter can be extracted from $e^+ e^-$ annihilations producing two back-to-back hadron pairs, and evolution equations are known to connect DiFFs at the different scales of the two reactions. 

In this paper, we have updated our first extraction of DiFFs from $e^+ e^-$ annihilation data in Ref.~\cite{Courtoy:2012ry} by performing the error analysis with the so-called replica method. The method is based on the random generation of a large number of replicas of the experimental points, in this case the Belle data for the process $e^+ e^- \to (\pi^+ \pi^-) \  (\pi^+ \pi^-)$~\cite{Vossen:2011fk}. Each replica is then separately fitted, producing an envelope of curves whose width is the generalization of the $1\sigma$ uncertainty band when the distribution is not necessarily a Gaussian. As such, this method allows for a more realistic estimate of the uncertainty on DiFFs. 

As a second step, we have used the above result to update our first extraction of the up and down valence transversities in a collinear framework~\cite{Bacchetta:2012ty}, employing data for two-particle-inclusive DIS off transversely polarized proton and deuteron targets. In particular, we have considered the recent measurement from the COMPASS collaboration for identified hadron pairs produced off transversely polarized proton targets~\cite{Braun:2015baa}. We have randomly generated replicas of these data and we have fitted them, making an error analysis similar to what has been done for DiFFs. We have noticed that many of these trajectories hit the Soffer bound, {\it i.e.} they lie close to the borders of the phase space where the $\chi^2$ function cannot be expected to have the quadratic dependence on the fit parameters as required by the standard Hessian method. Hence, we stress again that the replica method allows for a more reliable error analysis and we believe that the results shown in this paper represent the currently most realistic estimate of the uncertainties on the transversity distribution. 

As in our previous extraction~\cite{Bacchetta:2012ty}, we have adopted different scenarios for the functional form, all subject to the Soffer bound. We have further explored the sensitivity to the theoretical uncertainty on $\Lambda_{\text{QCD}}$ by using two different prescriptions for $\alpha_s (M_Z^2)$~\cite{Gluck:1998xa,Martin:2009iq}. In the range of experimental data, our results show little sensitivity to the variation of these parameters. The 68\% band for the valence up transversity turns out to be narrower than in the previous extraction because of the more precise COMPASS data on the proton target. Nevertheless, there is a significant overlap with the other existing parametrizations based on the Collins effect in single-hadron-inclusive DIS~\cite{Anselmino:2013vqa,Kang:2014zza}. The only source of discrepancy lies in the range $x \gtrsim 0.1$ for the valence down quark, where all replicas are driven to hit the lower Soffer bound irrespectively of the functional form and evolution parameters adopted. This behavior is induced by two specific bins in the set of COMPASS experimental data for the deuteron target. Since this data set is not changed with respect to our previous extraction, the present results just confirm those findings in Ref.~\cite{Bacchetta:2012ty}. It is interesting to note that also the down transversity of Ref.~\cite{Kang:2014zza}, extracted from the Collins mechanism but with evolution effects described in the TMD framework, tends to saturate the Soffer bound at $x > 0.2$. 

We have also calculated the first Mellin moment of transversity, {\it i.e.} the tensor charge, either by computing the integral upon the range of experimental data (truncated tensor charge) or by extrapolating the transversity to the full support $[0,1]$ in the parton fractional momentum $x$. The latter option obviously  induces a much larger error that somewhat decreases the relevance of the observed compatibility with the results obtained from the extraction based on the Collins effect. Nevertheless, we find good agreement also for the truncated tensor charges obtained in Ref.~\cite{Kang:2014zza}. We have also computed the isovector tensor charge $g_T$. 
The determination of the latter may shed light on hypothetical new elementary electroweak currents that are being explored through neutron $\beta$ decays~\cite{Bhattacharya:2011qm,Ivanov:2012qe,Cirigliano:2013xha,Courtoy:2014xea}, or even on direct dark matter searches~\cite{DelNobile:2013sia}. Our result has a very large error because, again, it requires the extrapolation of transversity outside the range of experimental data. Anyway, it is compatible with all the lattice results available in the literature. 

The large uncertainty caused by extrapolating the transversity reflects the need of two-particle-inclusive DIS data either at large and at very small $x$. More data have been released by the COMPASS collaboration that include also different types of hadron pairs (e.g., $K \pi$)~\cite{Braun:2015baa} and should allow to improve the flavour separation of transversity. More insight along the same direction will come also from polarized proton-proton collisions~\cite{Bacchetta:2004it}, where data for the semi-inclusive production of hadron pairs are expected from the PHENIX and STAR collaborations (see, e.g., Ref.~\cite{Vossen:2015jaa}). Finally, two-particle-inclusive DIS will be measured also at JLab in the near future, which should considerably increase our knowledge of transversity at large $x$.


\section*{Acknowledgments}
A.~Courtoy is funded by the Belgian Fund F.R.S.-FNRS via the contract of Charg\'ee de recherches. 
 


\bibliographystyle{jhep}
\bibliography{mybiblio}

\end{document}